\documentclass[11pt, fleqn]{paper}

\usepackage[utf8]{inputenc}
\usepackage[T1]{fontenc}
\usepackage{authblk}

\usepackage[colorlinks=true, citecolor=blue]{hyperref}

\usepackage[a4paper,margin=2.5cm, left=2cm, right=4cm ,marginparwidth=3cm]{geometry}	
\usepackage[onehalfspacing]{setspace}
\usepackage{lmodern}
\usepackage[dvipsnames]{xcolor}	
\usepackage{colortbl} %Shading to highlight in array
\usepackage{booktabs, multirow}
\newcolumntype{g}{>{\columncolor{lightgray}}c}

\usepackage{cite}
\usepackage[sort&compress,numbers]{natbib}

\usepackage{mathtools,amsmath, amssymb,amsthm}
\usepackage{tensor,mathrsfs,upgreek,stmaryrd}

\usepackage{xfrac}
\def\half{\tfrac{1}{2}}

\DeclareFontFamily{OT1}{rsfs}{}
\DeclareFontShape{OT1}{rsfs}{CGNPm}{n}{ <-7> rsfs5 <7-10> rsfs7 <10-> rsfs10}{}
\DeclareMathAlphabet{\mycal}{OT1}{rsfs}{CGNPm}{n}

{\catcode `\@=11 \global\let\AddToReset=\@addtoreset}
\AddToReset{equation}{section}

{\catcode `\@=11 \global\let\AddToReset=\@addtoreset}
\AddToReset{figure}{section}

{\catcode `\@=11 \global\let\AddToReset=\@addtoreset}
\AddToReset{table}{section}

\newcommand{\phere}{\bar\P}
\newcommand{\otherA}{\mathcal{A}}

\let\t\tensor
\let\p\partial
\def\dd{\mathrm{d}}

\let\t\tensor
\let\p\partial

\def\dd{\mathrm{d}}
\def\half{\tfrac{1}{2}}

\def\grav{\mathtt{g}} % gravitational acceleration
\def\poisson{\nu} % Poisson's ratio -- \nu and \sigma are both common
\newcommand{\shearmod}{\mu} % shear mudulus
\def\Dconst{ c^*} % gravitational acceleration

\def\R{\mathbb R}

\newcommand{\alphat}{\alpha}

\newcommand{\alpham}{\kappa}

\newcommand{\lapl}{\Delta_\delta}
\newcommand{\lapltwo}{\Delta^2_\delta}

\newcommand{\fsPhi}{\phi}
\newcommand{\FT}{F_T}
\newcommand{\FHertz}{\mathbb{F}}
\newcommand{\sigmaSi}{\sigma^{I}}
\newcommand{\sigmaSq}{\sigma}

\usepackage{textgreek}

\renewcommand{\P}{ \mathcal{P}}
\newcommand{\T}{\Theta}

\usepackage{subcaption}
\usepackage{float}
\usepackage{orcidlink}

\begin{document}
\title{On elastic deformations of cylindrical bodies under the influence of the gravitational field%
\protect\thanks{Preprint UWThPh-2024-3}
}

\author[1]{H.~Barzegar \orcidlink{0000-0002-6472-3115}}
\author[2]{P.~T.~Chruściel \orcidlink{0000-0001-8362-7340}}
\author[2,3]{F.~Steininger \orcidlink{0000-0003-0092-3043}}
\affil[1]{\footnotesize Universite Claude Bernard Lyon 1, CRAL UMR5574, ENS de Lyon, CNRS, Villeurbanne, F-69622, France}
\affil[2]{\footnotesize University of Vienna, Faculty of Physics and Research platform TURIS, Boltzmanngasse 5, 1090 Vienna, Austria}
\affil[3]{\footnotesize University of Vienna, Faculty of Physics, Vienna Doctoral School in Physics, Boltzmanngasse 5, 1090 Vienna, Austria}

\date{\today}
\maketitle

\begin{abstract}
We analyse the deformations of a cylindrical elastic body resulting from displacements in a varying gravitational field.
\end{abstract}
\tableofcontents

\section{Introduction}\label{sec:intro}

An experiment  is currently being built~\cite{HMMMCW} with the aim to measure the effect of the gravitational field on entangled states of photons propagating in an optical fiber.
\begin{figure}
  \centering
  \includegraphics[width=.5\textwidth]{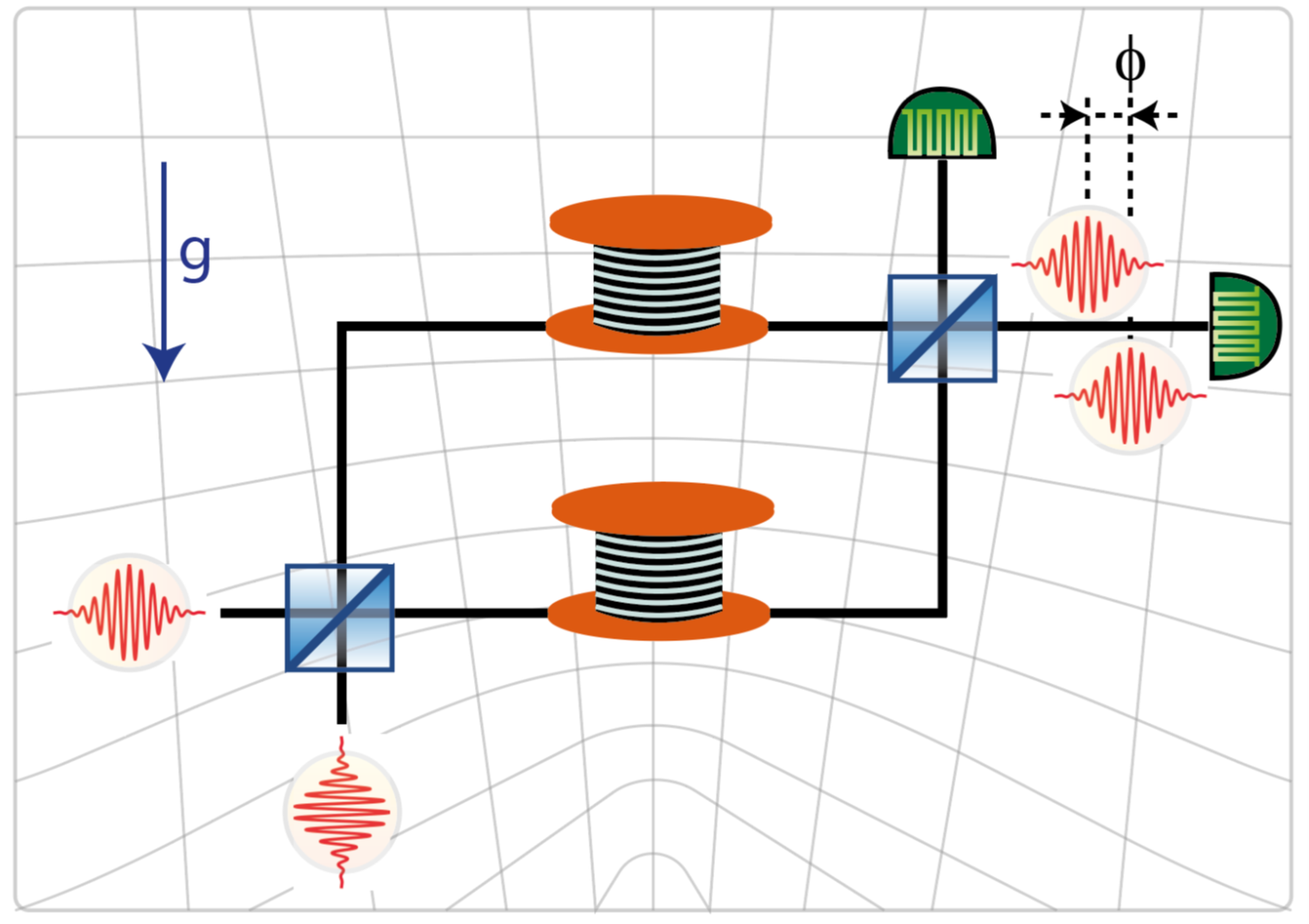}
  \caption{The GRAVITES experiment. The upper arm of the interferometer is moved vertically. The change of the gravitational field due to the change of height affects the propagation of light, resulting in a height-dependent phase shift.
Here and unless otherwise indicated, the $y$-axis is aligned vertically, so that the gravitational force acts anti-colinearly along the $y$-axis.
 \copyright~C.~Hilweg, reproduced with kind permission of the author. }\label{F17XI23.1}
\end{figure}
The experiment requires displacing vertically an optical fiber, with circular cross-sections, in a spooled configuration, see Figure~\ref{F17XI23.1}.
 Such a displacement in the gravitational field of the Earth leads to a minute phaseshift, which is expected to be measurable with the current state-of-the art photonic technology. The displacement is associated with a change of the ambient gravitational acceleration,
of temperature,
and of atmospheric pressure, leading to an elastic deformation of the fiber. The deformation affects  the shape, the length, and the
propagation properties of light in the fiber, leading to an additional phaseshift which needs to be determined for a correct interpretation of the results of the experiment.
The objective of this work is to develop a framework which can be used to determine this elastic deformation.
The resulting effects on the propagation properties of light in the fiber will be determined elsewhere.

\begin{figure}[htb]
	\centering
	\begin{subfigure}[t]{.4\textwidth}
		\centering
		\includegraphics[scale=1]{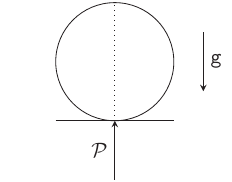}
		\caption{Resting on a contact line}
		\label{fig:one contact}
	\end{subfigure}
	\begin{subfigure}[t]{.4\textwidth}
		\centering
		\includegraphics[scale=1]{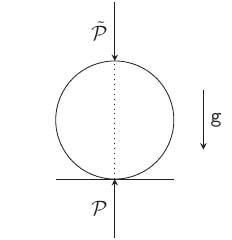}
		\caption{Squeezed between two lines}
		\label{fig:squeezed}
	\end{subfigure}
	\begin{subfigure}[t]{0.4\textwidth}
		\centering
		\includegraphics[scale=0.86]{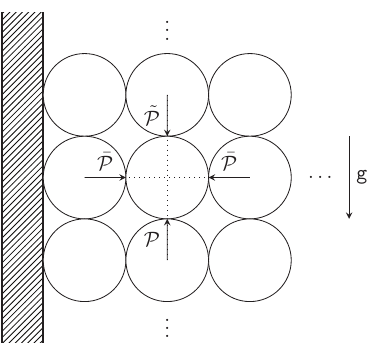}
		\caption{Quadrilateral configuration}
		\label{fig:coilquad}
	\end{subfigure}
	% \hfill
	\hspace*{-1.7em}
	\begin{subfigure}[t]{0.4\textwidth}
		\centering
		\includegraphics[scale=.9]{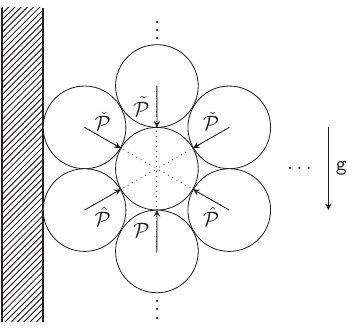}
		\caption{Hexagonal configuration}
		\label{fig:coilhexa}
	\end{subfigure}
	\caption{The four main models.   The $y$-axis is again along the vertical in all four figures. Figures (a) and (b) show  the cross-section of an infinite cylinder resting on a rigid support, with a supplementary pressure from the top in Figure (b). Figures (c) and (d) show    the spool of Figure~\ref{F17XI23.1} as cut by a vertical plane; we assume that  the radius of the spool is very large compared to the radius of the waveguide. The circles represent the consecutive returns of the waveguide, exercising pressure on the neighboring strands.   The dashed region represents the rigid spool.}
\end{figure}

The planned configuration is that of a spool with its axis of symmetry aligned vertically.
Since the radius of the spool is very large compared to the radius of the fiber, we ignore the spooling and consider a very long elastic cylinder. We then consider several models for the  problem at hand:

As a first model we start with a cylindrical waveguide resting on a horizontal contact line (see Figure~\ref{fig:one contact} and Section \ref{sec: one contact}).
The next model is a configuration where the  waveguide  is squeezed between two contact lines, to take into account the pressure arising from the  layers pressing from above  (see Figure~\ref{fig:squeezed} and Section \ref{sec: squeezed}).
The results obtained in both cases are  unacceptable throughout the waveguide,
with an infinite deformation at each contact line; this is of course a well known problem of such models (cf. \cite[Chapter 8.4.7]{Sadd7}). However, the model appears useful for its simplicity, we will return to this shortly.

To avoid the above problems we pass
to a Hertz-contact-type calculation, where we first analyse a configuration with the waveguide resting on a rigid support, followed by one where the waveguides are stacked upon each other;
in the last case the influence of the upper layer on a lower one is modelled by contact with a rigid plane. There  arises a  parameter describing the pressure from the waveguides stacked above the section of the fiber under consideration. We leave this parameter free in our calculations, its  value can be determined by  the number of layers and windings of the spool whenever a specific configuration is considered.

The above takes into account the vertical neighbours of any given section of the fiber, but ignores the side neighbours. To address this
we consider
two further configurations, where each fiber has four neighbours as in  Figure~\ref{fig:coilquad} (see Section \ref{sec: quadrilateral}),
or each fiber has six neighbours as in Figure \ref{fig:coilhexa} and Section \ref{sec: hexagonal}. The last model seems to us to provide the best approximation to the problem at hand. In such models new parameters arise, associated with  presence of new neighbours. We leave these parameters free again, and show in Appendix~\ref{App24XI23.1} how one of these parameters can be determined for each layer in the quadrilateral-contact case.

Our results show that in all models the  relative deformations are similar, for the practical purposes of our interest, close enough to the center of waveguide, where the guiding core resides. Therefore we expect that the first, simplest model, will be sufficient for the applications we have in mind.
We plan to return to this question in a future full treatment of the influence of small elastic deformations on the dispersion relation in optical fibers.

We include a constant ambient pressure term, as well as temperature effects  in all configurations in order to model possible variations of the environment.

\section{Waveguide Deformation in Earth's Gravity}\label{sec:Elasticity}

The unperturbed waveguide is modelled as a very long, homogeneous cylinder with radius $a$, length $L$ and density $\rho$, supported by a rigid plane with gravity acting as a body force via a constant gravitational acceleration $\grav$. We take $U = \{(r,\theta, z) : 0 < r\leq a, - \pi < \theta \leq \pi, 0 \leq z \leq L \}$ to be the interior of the waveguide and write $\p U$ for the surface $r = a$.
Note that $a \ll L$.

It is known that in isotropic and homogeneous materials the generalized Hooke's law in three dimensions yields the following stress-strain relation (cf., e.g., \cite[Eq.~(4.2.7)]{Sadd7} and \cite[Eq.~(4.6)]{Landau86}) in an  orthonormal frame, which we assume throughout,
\begin{equation}\label{eq: Hook}
	\sigma_{i j}
	=
	\lambda \epsilon_{k k} \delta_{i j}
	+
	2 \mu \epsilon_{i j}
\,,
\end{equation}
where $\lambda$ and $\mu$ are Lam\'e's first and second parameters of the material, respectively,  with $\mu$ being called the shear modulus which is sometimes  denoted by $G$ in the literature;
repeated indices are summed over unless explicitly indicated otherwise.
With $\sigma_{k k} = ( 3 \lambda + 2 \mu) \epsilon_{k k}$ we find
\begin{equation}\label{27XI23.8}
	\epsilon_{i j}
	=
	\frac{1 + \nu}{E} \sigma_{i j}
	-
	\frac{\nu}{E} \sigma_{k k} \delta_{i j}
\,,
\end{equation}
where $E := \mu (3 \lambda + 2 \mu)/(\lambda + \mu)$ is Young's modulus and $\nu := \lambda /[2(\lambda + \mu)]$ is  Poisson's ratio. Note that $E = 2 \shearmod (1 + \poisson)$.

When  the changes of temperature are \emph{not} negligible, the above needs to be revised as follows:
According to \cite[Section 4.4]{Sadd7},  in a thermally-isotropic and thermally-linear medium,
Equation~\eqref{27XI23.8} should be replaced by
\begin{equation}\label{eq: thermo-stress-strain}
    \epsilon_{i j}
	=
	\frac{1 + \nu}{E} \sigma_{i j}
	-
	\frac{\nu}{E} \sigma_{k k} \delta_{i j}
    +
   \alphat (T - T_0)  \delta_{ij}
\,,
\end{equation}
where $\alphat$ is the \emph{coefficient of linear thermal expansion}.
This is a consequence of the assumptions of linear thermo-elasticity, for which the strain decomposes into independent thermal and elastic components.
Inverting this relation, the stress tensor is given by 
\begin{equation}
    \sigma_{i j}
	=
	\lambda \epsilon_{k k} \delta_{i j}
	+
	2 \mu \epsilon_{i j}
    -
    (3 \lambda + 2 \mu)\alphat (T - T_0) \delta_{ij}
\,.
\end{equation}
\begin{figure}[htb]
	\centering
\begin{subfigure}{.45\textwidth}
    \centering
    \includegraphics[scale=.7]{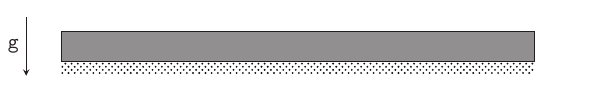}	
  \caption{Both ends open.}
	\end{subfigure}
	\qquad
  \begin{subfigure}{.45\textwidth}
		\centering
    \includegraphics[scale=.7]{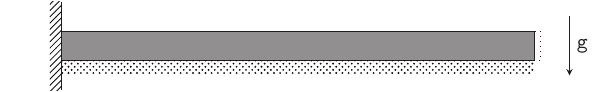}
	\caption{Fixed and open ends.}
	\end{subfigure}
\begin{subfigure}{.45\textwidth}
		\centering
    \includegraphics[scale=.7]{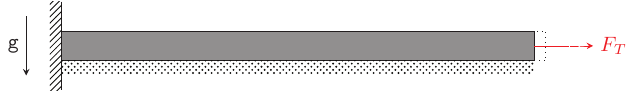}
  \caption{One fixed end and one under tension force $\FT$.}
	\end{subfigure}
	\qquad
\begin{subfigure}{.45\textwidth}
		\centering
    \includegraphics[scale=.7]{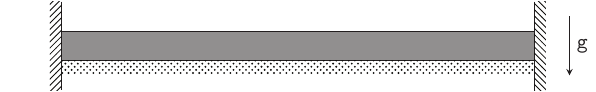}
  \caption{Both ends fixed. \\ \phantom{x}}
	\end{subfigure}
	\caption{Four models for a very long cylinder with boundary conditions covered by our calculations.
Vertical dashed regions represent fixed ends in the $z$-direction, whereas the dotted bars represent   rigid supports. The remaining boundaries in the figures can move freely.
 Note that  the model (d)  requires  the vanishing of the elongation coefficient $\alpham$ of \eqref{3VII23.1}.}
  \label{fig: idealized model}
\end{figure}

We allow the waveguide to stretch in the $z$ direction linearly in $z$  (compare Figure~\ref{fig: idealized model} (a)-(c)):
\begin{equation}\label{3VII23.1}
  u_z = \alpham  z
  \,.
\end{equation}
All remaining fields are assumed to be $z$-independent.
Equation~\eqref{3VII23.1} leads to a non-vanishing $z $-component of the strain tensor
\begin{equation}
\epsilon_{ij} =
 {1\over 2}
  ( \partial_i u_j +\partial_j u_i)
  \,,
\end{equation}
%:
namely
\begin{equation}\label{3VII23.2}
   \epsilon_{zz} = \alpham\,, \ \mbox{but}
   \
   \epsilon_{xz}=\epsilon_{yz}=0
    \,,
\end{equation}
giving an additional contribution to the usual plane-strain relations
(cf. \cite{Zhenye90} for related considerations).

Having reduced the dimensionality of the problem, the equilibrium equations  (cf., e.g., \cite[Eq.~(3.6.4)]{Sadd7} or \cite[Eq.~(2.34)]{Gould13}),
\begin{equation}
	\p_j \t{\sigma}{_i_j} + F_i= 0 \,,
\end{equation}
with $\t{\sigma}{_i_j}$ the stress tensor and $F_i = -\p_i V$ the body force due to gravity, where $V = \grav \rho y$, simplify to
\begin{align}
	\p_x \t{\sigma}{_x_x} + \p_y \t{\sigma}{_x_y} - \p_x V
		& = 0  \,,
	\\
	\p_x \t{\sigma}{_x_y} + \p_y \t{\sigma}{_y_y} - \p_y V
		& = 0 \,.
\end{align}

Note that the stress in the $z$ direction does not necessarily vanish. Indeed, the $zz$-component of \eqref{eq: thermo-stress-strain} gives
\begin{equation}
	\alpham
	=
	\frac{\sigma_{zz}}{E}
	-
	\frac{\poisson}{E} (\sigma_{rr} + \sigma_{\theta \theta})
	+
	\alphat (T - T_0)
\,,
\end{equation}
so that
\begin{equation}
\t{\sigma}{_z_z} =
   \nu (\t{\sigma}{_x_x} + \t{\sigma}{_y_y})
+ E  \alpham
	-
	E \alphat (T - T_0)
 \,.
  \label{21XI23.1}
\end{equation}

Adapting our coordinate system to the symmetry of the setup by choosing cylindrical coordinates,
from now on tensor components will  refer to the following orthonormal frame, which in Cartesian coordinates is given by
\begin{equation}
	e_r : = \big( \cos (\theta), \sin (\theta), 0\big) \,, \quad  e_\theta : = \big( -\sin (\theta), \cos (\theta), 0\big) \,, \quad  e_z : = ( 0,0,1) \,.
\end{equation}
In this frame the stress tensor can be expressed in terms of derivatives of the Airy stress function $\phi$ as
(cf. \cite{Airy} and, e.g., \cite[Eqs.~(8.12)--(8.13)]{Barber})
\begin{align}\label{eq:stress_cyl_rr}
	\t{\sigma}{_r_r}
		&
  = \tfrac{1}{r} \p_r \phi + \tfrac{1}{r^2} \p_\theta^2 \phi + V
   \,,
	\\
	\t{\sigma}{_\theta_\theta}
		& = \p_r^2 \phi + V
   \,,
		\label{eq:stress_cyl_tt}
	\\
	\t{\sigma}{_r_\theta}
		& = -\p_r (\tfrac{1}{r} \p_\theta \phi) \,,
		\label{eq:stress_cyl_rt}
\end{align}
which has to satisfy the compatibility condition
(see, e.g., \cite[Eq.~(7.5.5)]{Sadd7} or \cite[Eq.~(7.17b)]{Gould13}, together with  \cite[Eq.~(12.3.7)]{Sadd7}),
\begin{equation}\label{eq:gravity:biharmonic}
	\lapltwo \phi = - \frac{1 - 2 \poisson}{1 - \poisson} \lapl V
	-
   \frac{E \alphat }{1-\nu} \lapl T
    \,,
\end{equation}
where $\lapl$ is the Laplace operator of the Euclidean metric on $\R^2$.
(Note that $V$ is only defined up to a constant, which can be absorbed by a redefinition of $\phi$.)
In this work we consider a  steady-state configuration, which requires $\lapl T = 0$ (cf.~\cite[Section 12.1]{Sadd7}). Further,
since $V= \grav  \rho y$ we have $\lapl V = 0$ as well, implying that the compatibility conditions reduce to the homogeneous biharmonic equation.

Since the frame $\{e_r,\, e_\theta, \, e_z\}$ is orthonormal, the strain is related to the stress via \eqref{eq: thermo-stress-strain}:
\begin{align}\label{eq:strain_cyl_rr}
	\t{\epsilon}{_r_r}
	& =
	\frac{1}{2 \shearmod}
	\left[
		(1 - \poisson) \t{\sigma}{_r_r} -  \poisson \t{\sigma}{_\theta_\theta}
	\right]
	-
	\poisson \alpham
	+
	 (1 + \poisson )\alphat (T -T_0)
	 \,,
\\
	\t{\epsilon}{_\theta_\theta}
	& =
	\frac{1}{2 \shearmod}
	\left[
		(1 - \poisson) \t{\sigma}{_\theta_\theta} - \poisson \t{\sigma}{_r_r}
	\right]
	-
	\poisson \alpham
	+
	 (1 + \poisson )\alphat (T -T_0)
\,,
\label{eq:strain_cyl_tt}
	\\
	\t{\epsilon}{_r_\theta}
		& = \tfrac{1}{2 \shearmod} \t{\sigma}{_r_\theta} \,.
		\label{eq:strain_cyl_rt}
\end{align}
One can now determine the displacement vector $u_i$ from  the usual equations, where $u_r$ and $u_\theta$ are frame components of $u$ (cf., e.g., \cite[Eq.~(7.6.1)]{Sadd7})
\begin{align}
\label{eq:StrainDisplacementrr}
	\t{\epsilon}{_r_r} &= \p_r u_r\,,
		\\
	\t{\epsilon}{_\theta_\theta} &= \tfrac{1}{r} \left( \p_\theta u_\theta + u_r \right)\,,
		\\
\label{eq:StrainDisplacementrt}
	\t{\epsilon}{_r_\theta} &= \half \left( \tfrac{1}{r} \p_\theta u_r + \p_r u_\theta - \tfrac{1}{r} u_\theta \right)\,.
\end{align}
\begin{figure}[tb]
  \centering
  \includegraphics[scale=1]{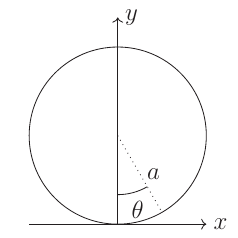}
  \caption{Convention  for polar coordinates.}
  \label{fig:waveguidemodel}
\end{figure}

Having compiled all relevant equations from linear elasticity, we set up an adapted coordinate system as in Figure \ref{fig:waveguidemodel}, with
\begin{equation}
	r \in (0, a] \ \mbox{and} \ \theta \in (- \pi , \pi ]\,.
\end{equation}

There exists a general solution for \eqref{eq:gravity:biharmonic} in polar coordinates, denoted by\footnote{Note that the redefinition $\theta \rightarrow \theta - \tfrac{\pi}{2}$ can be absorbed by the constants.}
\begin{align}
	 \fsPhi (r,\theta)=
\notag	& A_0 \ln r + B_0 + C_0 r^2 \ln r + D_0 r^2
		\\
\notag	&+ \left( a_0 \ln r + b_0 + c_0 r^2 \ln r + d_0 r^2 \right) \theta
		\\
		&+ \left( A_1 r \ln r + \tfrac{B_1}{r} + C_1 r^3 + D_1r + E_1 r \theta + F_1 r \theta \ln r \right) \cos (\theta)
 \notag	
		\\
\notag	&+ \left( a_1 r \ln r + \frac{b_1}{r} + c_1 r^3 + d_1 r + e_1 r \theta + f_1 r \theta \ln r \right) \sin (\theta)
		\\
\notag	&+ \sum_{n\geq 2} \Big[ \left( A_n r^n + B_n r^{-n} + C_n r^{n + 2} + D_n r^{2-n} \right) \cos(n \theta)
		\\
&\qquad+ \left( a_n r^n + b_n r^{-n} + c_n r^{n + 2} + d_n r^{2 - n}\right) \sin (n \theta)\Big]\,,
 \label{eq:Michell}
\end{align}
known as the `Michell solution' \cite{Michell}.

In practical terms, we are mostly concerned with determining the coefficients in \eqref{eq:Michell} from the boundary conditions we impose. An immediate simplification can be achieved by only considering solutions which have  the mirror symmetry $x \mapsto - x$ (equivalently $\theta \mapsto - \theta$), since the forces  considered are invariant under this symmetry;
see Appendix~\ref{app:non-symmetric} for a treatment without imposing mirror symmetry at the outset.
Additionally, we require regularity at $r=0$ and $2\pi$-periodicity in $\theta$. Restricting to mirror-symmetric boundary conditions finally lead to a requirement of mirror-symmetry in $\t{\sigma}{_r_r}$ and $\t{\sigma}{_\theta_\theta}$ and anti-symmetry for $\t{\sigma}{_r_\theta}$.
Lastly, note that the parameters $B_0$ and $D_1$ do not contribute to the stress as by Eqs.~\eqref{eq:stress_cyl_rr}--\eqref{eq:stress_cyl_rt}, which is straightforwardly verified. They may be considered degeneracies of the solution space and set to zero without loss of generality (cf. \cite{Barber}).
The remaining terms in \eqref{eq:Michell} are
\begin{align}\label{eq:Michell_adapted}
	 \fsPhi (r,\theta)=&  D_0 r^2
		+ C_1 r^3 \cos (\theta)
		+ \sum_{n\geq 2} \left( A_n r^n +  C_n r^{n + 2} \right) \cos(n \theta)\,,
\end{align}
which is the form considered for all applications below.

The displacement can be calculated via \eqref{eq:StrainDisplacementrr}--\eqref{eq:StrainDisplacementrt}\footnote{A convenient form for the full parametrized solution can be found in \cite[Tab.~(9.1)]{Barber}.}. We have for the remaining terms in $\fsPhi$,
\begin{align}
\notag	
	u_r (r, \theta)
	=\,&
	\frac{1}{2 \shearmod}
		\Big\{ 2 (1 - 2 \nu) D_0 r + (1 - 4 \nu) C_1 r^2 \cos (\theta)
		-
		\half \grav \rho (1 - 2 \nu) r^2 \cos (\theta)
\\
		&+
		\sum_{n \geq 2} \left[- n A_n r^{n-1} + (2 - 4\nu - n) C_n r^{n+1} \right] \cos (n \theta)
		\Big\}
		 -
	\Xi \cos (\theta)
	-
	\poisson \alpham  r
 \nonumber
\\
	&
	+
	(1 + \poisson )\alphat (T -T_0) r
\,,
	\label{eq:displacement_admissible_ur}
\end{align}
\begin{align}
	u_\theta (r, \theta)
\notag		
	=\,&
	\frac{1}{2 \shearmod}
	\Big\{
		(5 - 4 \nu) C_1 r^2 \sin (\theta)
		-
		\half \grav \rho (1 - 2 \nu) r^2 \sin (\theta)
\\
	& +
	\sum_{n \geq 2} \left[ n A_n r^{n-1} + (4 - 4\nu + n) C_n r^{n+1} \right] \sin (n \theta)
	\Big\}
+
	\Xi \sin (\theta)
	+ \Dconst r
\,,
\label{eq:displacement_admissible_ut}
\end{align}
where $\Dconst$, $\Xi$ are integration constants.
These are fixed by the boundary conditions imposed at the contact point $u_r (a ,  0) = 0 = u_\theta (a , 0) $, which imply $\Dconst = 0$ and
\begin{equation}
\begin{aligned}
	\Xi
	&:=
	\frac{1}{2 \shearmod}
	\Big\{
		2 (1 - 2 \nu) D_0 a + (1 - 4 \nu) C_1 a^2
		-
		\half \grav \rho (1 - 2 \nu) a^2
\\
& \qquad\quad
		+
		\sum_{n \geq 2} \left[- n A_n a^{n-1} + (2 - 4\nu - n) C_n a^{n+1} \right]
	\Big\}
	-
	\poisson \alpham  a
	+
	(1 + \poisson )\alphat (T -T_0) a
\,,
\end{aligned}
 \label{24XI23.p1}
\end{equation}
provided that the sums converge.

\section{Boundary Conditions}

We now have general expressions for the stresses \eqref{eq:stress_cyl_rr}--\eqref{eq:stress_cyl_rt},
 the strains \eqref{eq:strain_cyl_rr}--\eqref{eq:strain_cyl_rt} and the displacements \eqref{eq:displacement_admissible_ur}--\eqref{eq:displacement_admissible_ut} in terms of the coefficients $\{ D_0, C_1, A_n, C_n\}$ for $n \geq 2$.
These coefficients can be determined from the boundary conditions algebraically.

At the boundary of the cylinder we consider an angle dependent pressure   $f(\theta)$ acting in the radial direction and a shear force per unit area  $g(\theta)$. The boundary conditions in linear elasticity require the stresses at the boundary to react to the external forces as
\begin{align}
	\t{\sigma}{_r_\theta}|_{\p U} &= g (\theta)\,,
		\\
	\t{\sigma}{_r_r}|_{\p U} &= f (\theta)\,.
 \label{29VI23.1}
\end{align}
Our assumption of mirror-symmetry implies that the boundary conditions can be Fourier decomposed into sine and cosine series respectively,
\begin{align}
	\t{\sigma}{_r_\theta}|_{\p U} &= \sum_{n \geq 0} g_n \sin (n \theta)\,,
		\\
	\t{\sigma}{_r_r}|_{\p U} &= \frac 12 f_0 + \sum_{n \geq 1} f_n \cos(n \theta)\,,
	\label{eq:cosexp}
\end{align}
with the coefficients given explicitly by
\begin{align}
	g_n = \frac{2}{\pi} \int_0^\pi \dd \theta\; g(\theta) \sin (n \theta)\,,
		\\
	f_n = \frac{2}{\pi} \int_0^\pi \dd \theta\; f(\theta) \cos (n \theta)\,.
\end{align}

On the other hand, we know that the solution can be written using \eqref{eq:stress_cyl_rr}--\eqref{eq:stress_cyl_rt} with \eqref{eq:Michell_adapted} for the Airy stress function; explicitly
\begin{align}
	\t{\sigma}{_r_\theta}|_{\p U} =\,& 2 C_1 a  \sin (\theta) + \sum_{n \geq 2} \left[ (n-1) A_n + a^2 (n+1) C_n \right] n a^{n-2} \sin (n \theta)\,,
		\\
	\t{\sigma}{_r_r}|_{\p U} =\,& 2 D_0 + (2 C_1 - \grav \rho) a \cos(\theta)
		\\
	\notag &- \sum_{n \geq 2} \left[ (n-1) n A_n + a^2 (n^2 - n -2) C_n \right] a^{n-2} \cos(n \theta)\,.
\end{align}

Comparing term by term  determines the coefficients.
We are generally interested in the case with vanishing shear forces,
 i.e. $g(\theta) = 0$, which corresponds to only normal forces or the frictionless limit. Then,
\begin{equation}
	C_1 = 0 \ \mbox{and} \ C_n = \frac{1 - n}{a^2 (1+n)} A_n\,,
\end{equation}
and for the radial stress
\begin{align}
	\t{\sigma}{_r_r}|_{\p U} &= 2 D_0  - a \grav \rho \cos(\theta) - 2 \sum_{n \geq 2} A_n (n-1)  a^{n-2} \cos(n \theta)\,.
\end{align}

Comparing coefficients with \eqref{eq:cosexp}, we find
\begin{align}
	D_0 &= \frac 14 f_0\,,
		\\
	A_n &= -\frac{1}{2 (n-1)} a^{2-n} f_n\,,
\end{align}
for $n\geq 2$, with
\begin{equation}
	f_1 = - a \grav \rho\,.
\end{equation}
We see that the Fourier coefficient $f_1$ in  the function $f$ in \eqref{29VI23.1} is not arbitrary, and is determined by the body force.

The boundary conditions in the $z$-direction are given by the models shown in Fig.~\ref{fig: idealized model}. We focus on subfigure (c), with the special case without tension force $\FT$ corresponding to subfigure (b).
By definition, the displacement boundary conditions are given by \eqref{3VII23.1}, since we allow for elongation of the open end.
The boundary condition for the stress is given by
\begin{equation}\label{eq:10X23.2} 
\FT = 
\int_{U(z=L)}\sigma_{zz} r \, \dd r \, \dd\theta
	\,,
\end{equation}
 i.e.\ the forces on the end face of the cylinder have to match the stresses on the end face.
Using \eqref{21XI23.1}, this can be rewritten as
\begin{equation}\label{eq:10X23.2bc}
\FT =    \int_{U(z=L)}
  \nu (\t{\sigma}{_x_x} + \t{\sigma}{_y_y})
   r \, \dd r \, \dd\theta   
   +  \pi a^2E
    \big[  \alpham
	-
	 \alphat (T - T_0)
 \big]
 \,.
\end{equation}
The integral can be calculated either explicitly or numerically for the solutions in Section~\ref{sec:contact_models}, 
 providing thus a relation between the forces acting on the system, $\alpham$, and the remaining parameters that appear in the problem.

This can be restated as an equation for the elongation $\alpham$ as a function of tension $\FT$ and the plane stresses along the fiber, so
\begin{equation}\label{eq:10X23.2bc}
\alpham =    - \frac{1}{\pi a^2E }\int_{U(z=L)}
  \nu (\t{\sigma}{_x_x} + \t{\sigma}{_y_y})
   r \, \dd r \, \dd\theta   +
	 \alphat (T - T_0)
 + \frac{\FT}{\pi a^2E }
\,.
\end{equation}
For a waveguide of total length $L$, this implies a change in length
\begin{equation}\label{eq:10X23.2}
	L \mapsto L+\alpham L  =
  L
  \Big[1 - \frac{1}{\pi a^2E }\int_{U(z=L)}
  \nu (\t{\sigma}{_x_x} + \t{\sigma}{_y_y})
   r \, \dd r \, \dd\theta   +
	 \alphat (T - T_0)
 + \frac{\FT}{\pi a^2E }
  \Big]
	\,.
\end{equation}

\section{Contact Models}
\label{sec:contact_models}

Having formally solved the problem for arbitrary boundary conditions in the preceding section, we now implement the   boundary conditions for the configurations shown in Figures~\ref{fig:one contact}--\ref{fig:coilhexa}.

As already pointed out in the Introduction, the solutions for the configurations of Figures~\ref{fig:one contact}--\ref{fig:squeezed}, that we are about to derive, with the boundary conditions corresponding to contact lines,  describe unphysical displacement fields, diverging at the contact interfaces. We show that  this can be cured  by deriving a solution involving extended contact regions, using the Hertz contact deformations formalism.
We show that even though the displacements differ between these approaches, the stresses near the center of the waveguide  are in reasonable agreement.

\subsection{Line contacts}
\subsubsection{Resting on a contact line}
\label{sec: one contact}

The simplest physical model is given by a cylindrical waveguide lying on an infinite plane, contacting as a first approximation only on a line.
This is reminiscent of the famous Flamant solution with circular cross-section (see in particular \cite[Problem 3 of Chapter 12]{Barber}).

The boundary conditions are given by
\begin{align}
	\t{\sigma}{_r_\theta}|_{\p U} &= 0\,,
		\\
	\t{\sigma}{_r_r}|_{\p U} &= \P \delta(\theta)
 - \mathfrak{p}\,,
\end{align}
where $\P$ is the pressure with which the contact line is resisting the weight of the waveguide and $\mathfrak{p}$ the ambient pressure, taken to be constant.
To put this into physical terms, we   apply neither tangential  nor radial forces, except for the contact line with the ``contact wire''.

This model is particularly convenient, since the $\delta$-distribution has a Fourier series expansion as
\begin{equation}
	\delta (\theta) = \frac{1}{2\pi} \left[1 + 2 \sum_{1 \leq n} \cos (n \theta) \right]\,.
\end{equation}
We find for the coefficients in the Michell solution
\begin{align}
	D_0 &= \frac{\P}{4\pi} - \frac{\mathfrak{p}}{2}\,,
	\label{29VII22.1a}
		\\
	C_1 &= 0\,,
		\\
	A_n &= -\frac{\P}{2\pi a^{n-2} (n-1)}\,,
		\\
	C_n &= \frac{\P}{2\pi a^n (n+1)}\,,
\label{29VII22.1d}
\end{align}
with
\begin{equation}
	\P = - \pi a \grav \rho\,.
\end{equation}

This last constraint implements the physicality of the calculations, since in the absence of other forces there can only be an equilibrium configuration if the integrated body force of the waveguide $\int (- F_y) r\, \dd r \dd \theta = \int \grav \rho\, r\, \dd r \dd \theta = \pi a^2 \grav \rho$
matches the reactive force exterted by the plane, given by $\int \P \delta(\theta) r \, \dd \theta |_{\p U} = - \pi a^2 \grav \rho$.

\begin{figure}[t]
  \centering
  \includegraphics[scale=0.4]{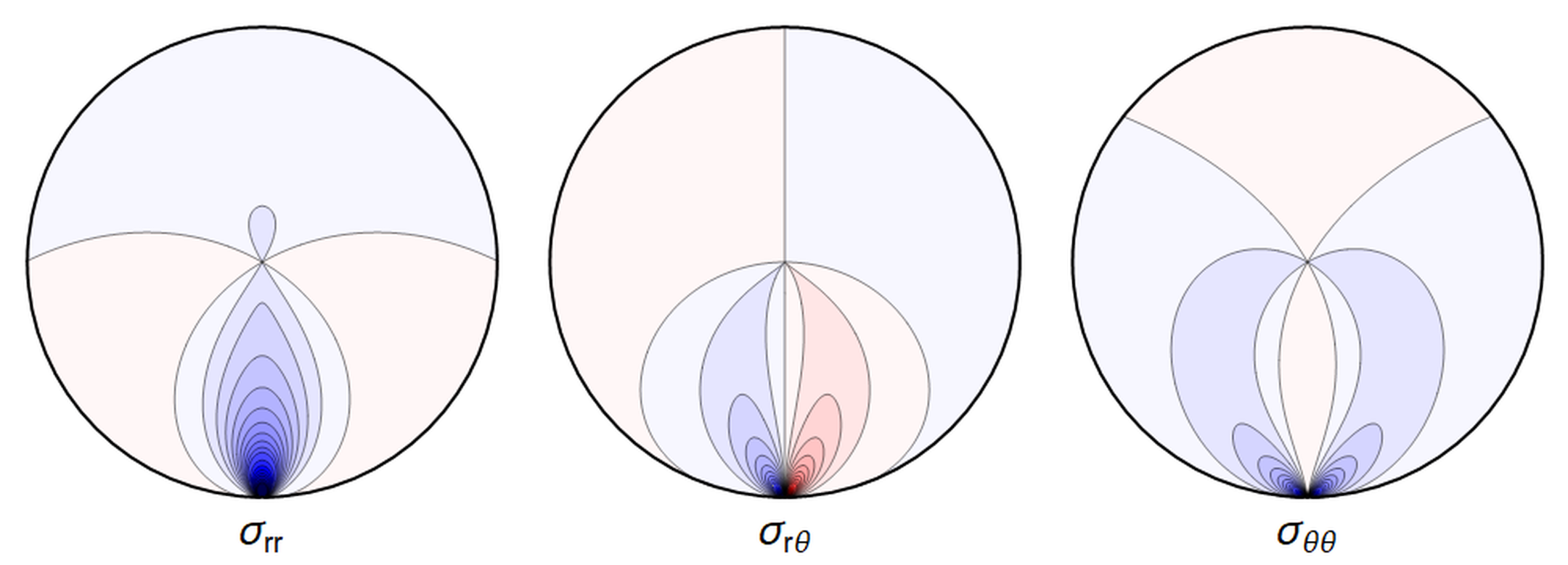}
  \includegraphics[scale=0.4]{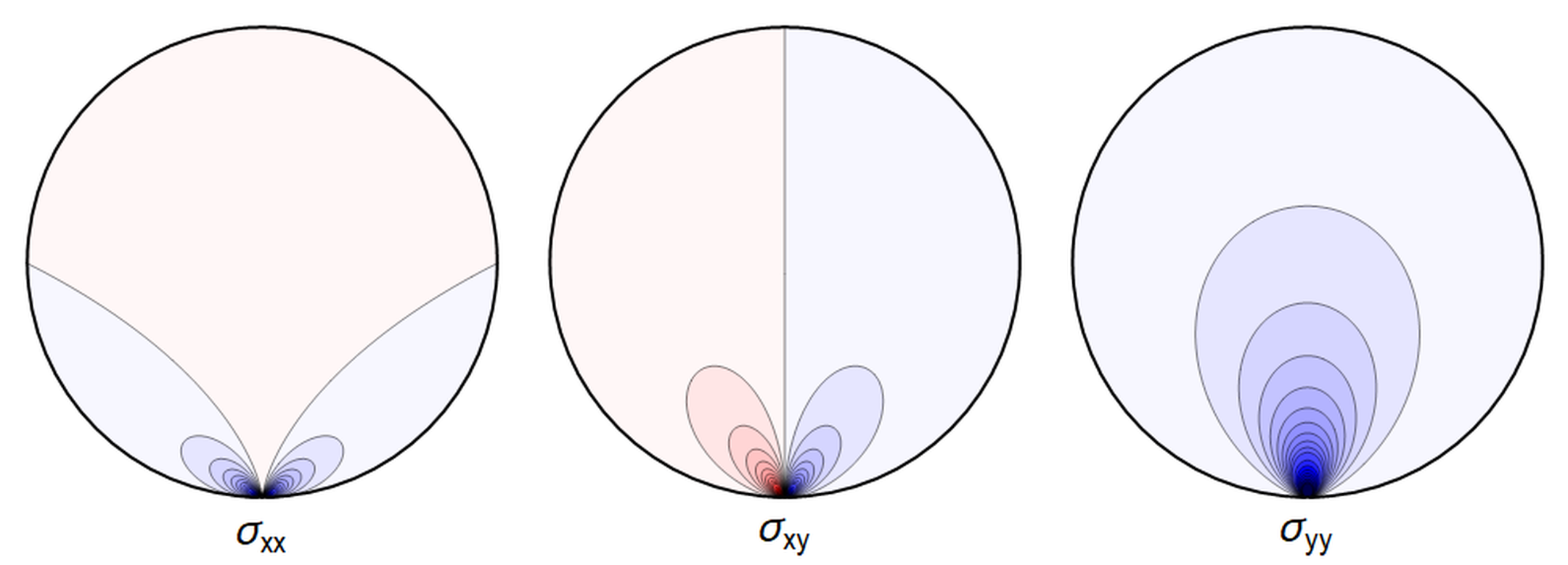}
  \caption{Typical  plot of internal stresses for the case of a waveguide resting on an infinitely thin line, with unrealistic parameters arbitrarily chosen for illustration purposes listed in Table~\ref{tab:visual_parameters} . Darker colors encode larger stresses, with blue color indicating compressive (negative) stresses.
  }
  \label{fig:stress_line}
\end{figure}

The sum in \eqref{eq:Michell_adapted} converges for coefficients \eqref{29VII22.1a}--\eqref{29VII22.1d}, yielding
\begin{align}
	 \fsPhi (r,\theta)=
		& - \frac 12 r^2 \mathfrak{p} + \frac 14  r \grav \rho \left[ r \left( a + r \cos (\theta) \right) - 4 a^2 \arctan \left( \tfrac{ r \sin (\theta) }{ a - r \cos (\theta)} \right) \sin (\theta) \right]
\,,
\label{29VII22.2}
\end{align}
and further for the stresses
\begin{align}
\notag
	\sigma_{rr} =\,&
		 A \left[ r (6 a^2 + r^2) \cos (\theta) - a \left( a^2 + 3 r^2 + 2 (a^2 + r^2) \cos (2 \theta) - a r \cos (3 \theta) \right) \right]
\\
   &-\mathfrak{p}
\,,
\label{29VII22.3a}
\\
	\sigma_{r\theta} =\,&
		A \left[ 4 a ( a^2 + r^2 ) \cos (\theta) - r \left( 5 a^2 + r^2 + 2 a^2 \cos (2 \theta)\right) \right] \sin (\theta)
\,,
\label{29VII22.3b}
\\\notag
	\sigma_{\theta\theta} =\,&
		A \left[ - r (2 a^2 + r^2) \cos (\theta) + a \left( r^2 - a^2 + 2 (a^2 + r^2) \cos (2 \theta) - a r \cos (3 \theta) \right) \right]
\\
  &-\mathfrak{p}
\,,
\label{29VII22.3c}
\end{align}
with
$$
 A := \frac{\grav \rho (a^2 - r^2)}{2 \left[a^2 + r^2 - 2 a r \cos (\theta) \right]^2}
 \,.
$$
%.
A representative plot of the stresses can be seen in Figure~\ref{fig:stress_line}.
\begin{table}[h!]
$$
\begin{array}{||c|c|c|c|c||}
\hline\hline
	\rho \grav	& \poisson	&	\mathfrak p\,, \alpham \,, \alphat		& \tilde \P & \hat \P = \check \P
	 \\\hline
	0.01	& 	0.17	&	0	 &	1 & 0.3
	 \\ \hline\hline
\end{array}
$$
\caption{Set of parameters chosen in our visualizations. For conciseness, we express lengths in multiples of $a$,   and pressure in multiples of the shear modulus $\mu$, leading to dimensionless quantities for the remaining parameters used in numerical calculations.}
\label{tab:visual_parameters}
\end{table}

\begin{figure}[h!]
  \begin{center}
  \includegraphics[scale=0.4]{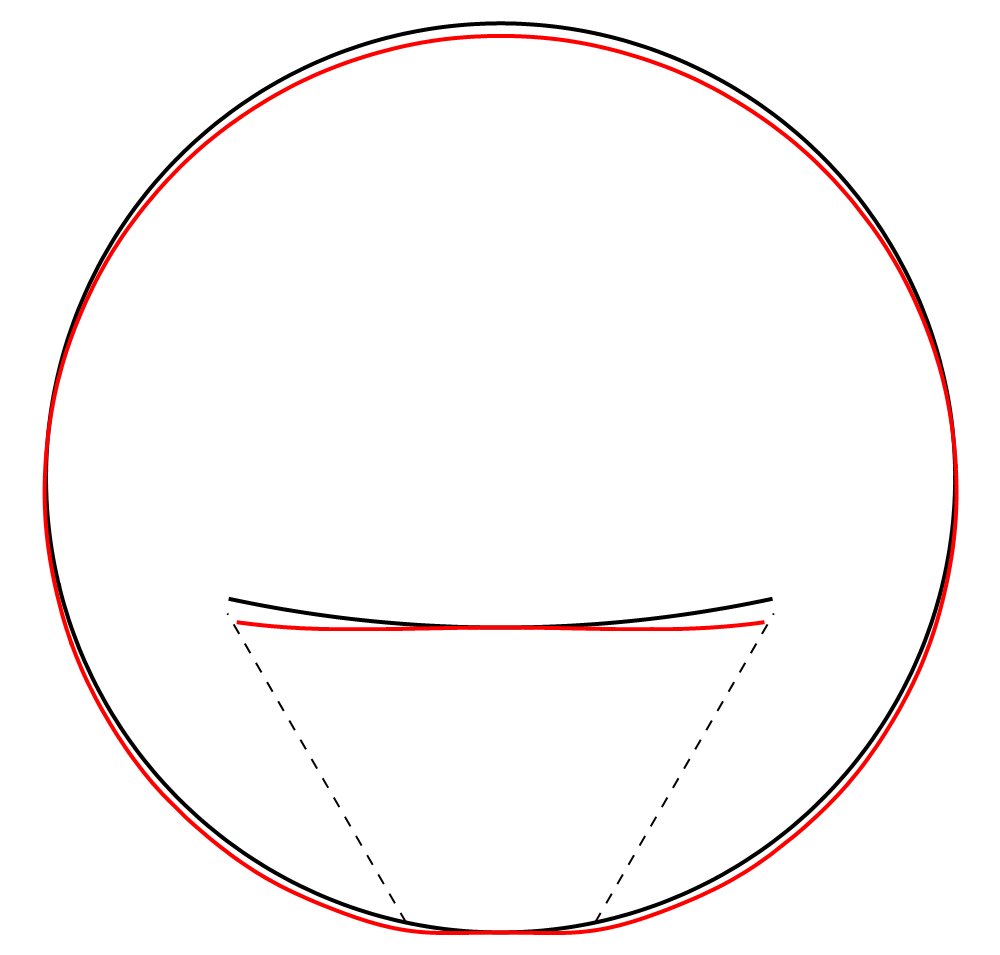}
  \end{center}
  \caption{Illustrative deformation for the case of a waveguide resting on an infinitely thin line, after truncating the sums in \eqref{eq:displacement_admissible_ur}-\eqref{24XI23.p1} to $n=20$, drawn in red. The undeformed reference is drawn in black. Here and in similar figures below, the inset shows a magnified section around the contact point.}
  \label{fig:deformation_line}
\end{figure}

Equations \eqref{eq:displacement_admissible_ur}-\eqref{24XI23.p1} do not make sense, as the sum in $\Xi$ is infinite. But one can find explicit expressions for $u_r$ and $u_\theta$ by integration; the resulting formulae are lengthy and not very enlightening, therefore we did not include them here.
Not unexpectedly, and similar to the Flamant solution \cite[Chapter 8.4.7]{Sadd7} ,
the singularity in the function $A$ at $r=a$ and $\cos \theta =1$
leads to an infinite displacement there. Hence the boundary condition $u(r=a,\theta=0)=0$ cannot be imposed. However, the displacements obtained by direct integration are finite away from the contact point, in particular near the center of the waveguide (cf.~\cite{1951_Timoshenko_Goodier_elasticity}).  One can also truncate the series \eqref{eq:displacement_admissible_ur}-\eqref{eq:displacement_admissible_ut} which
leads to finite solutions, illustrated in Figure~\ref{fig:deformation_line}.

\subsubsection{Squeezed between two lines}
\label{sec: squeezed}

A second straightforward case is given by the waveguide being squeezed between two lines, with an arbitrary pressure $\tilde \P$ pushing from the top.
See also \cite[Example 8-10]{Sadd7} and \cite[Problem 1 of Chapter 12]{Barber} where a similar problem is modelled by superposition of three particular stress fields including two Flamant solutions together with a uniform radial tension loading.
The boundary conditions are given by
\begin{align}
	\t{\sigma}{_r_\theta}|_{\p U} &= 0\,,
		\\
	\t{\sigma}{_r_r}|_{\p U} &= \P \delta(\theta) + \tilde \P \delta (\theta - \pi )  - \mathfrak{p}\,.
\end{align}
The calculation is completely analogous to the above, with the boundary condition enforcing
\begin{align}
	D_0 &= \frac{\P + \tilde \P}{4\pi}  - \frac{\mathfrak{p}}{2}\,,
		\\
	C_1 &= 0\,,
		\\
	A_n &= -\frac{\P + (-1)^{n} \tilde \P}{2\pi a^{n-2} (n-1)}\,,
		\\
	C_n &= \frac{\P + (-1)^{n} \tilde \P}{2\pi a^n (n+1)}\,,
\end{align}
with
\begin{equation}
	\P = \tilde \P - \pi a \grav \rho\,.
\end{equation}
Again, the physical interpretation is that now the force from above implies a larger necessary reactive force from below to achieve equilibrium.

\begin{figure}[t]
  \centering
  \includegraphics[scale=0.4]{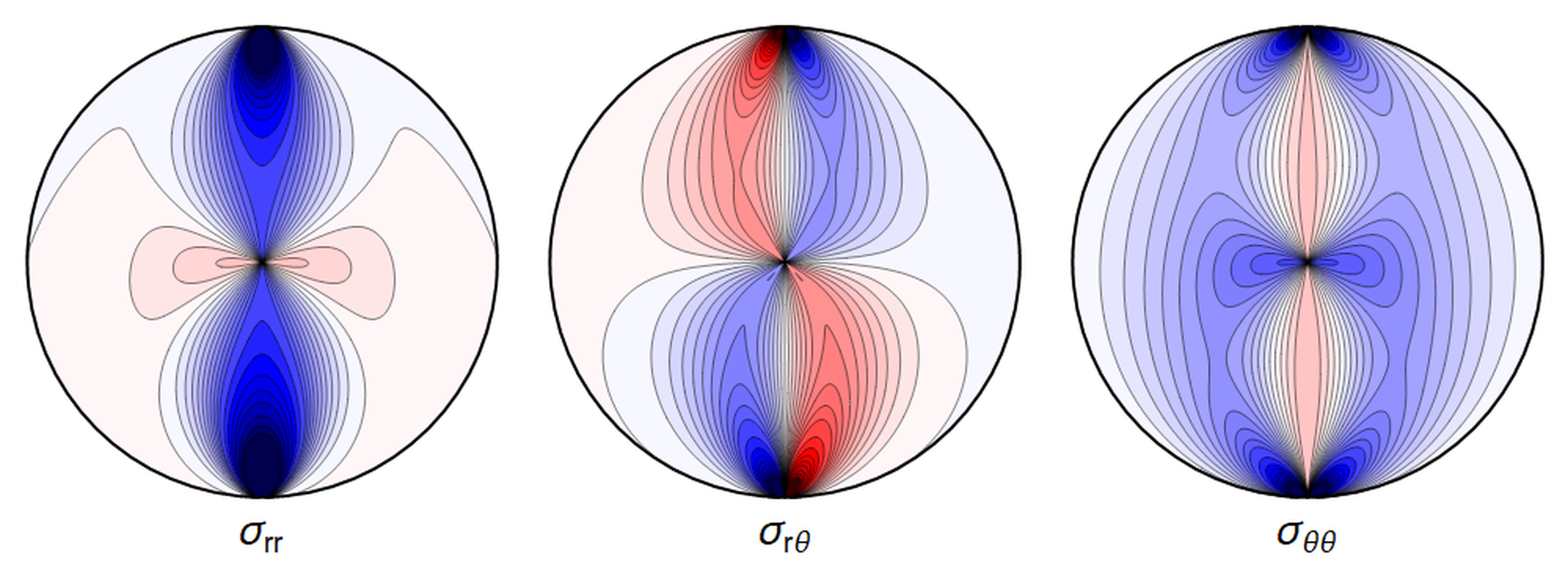}
  \includegraphics[scale=0.4]{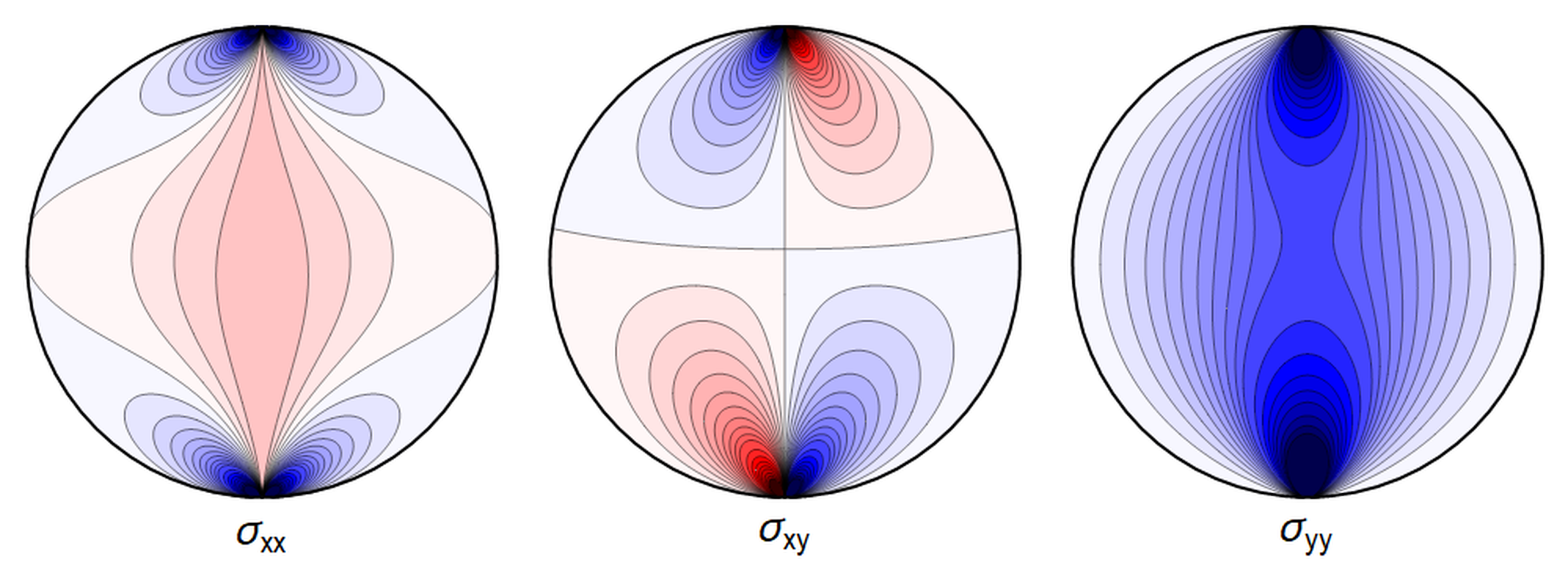}
  \caption{Typical plot of internal stresses for the case of a waveguide squeezed between two lines; same  parameters and color coding as in Fig.~\ref{fig:stress_line}.}
  \label{fig:stress_squeeze}
\end{figure}

Denoting by $\sigmaSi_{ij}$ the right-hand sides of \eqref{29VII22.3a}--\eqref{29VII22.3c}, the stress components read
\begin{align}
	\sigmaSq_{rr} &=
		 \sigmaSi_{rr} - B ( a^2 - r^2 ) \left[ a^4 - 2 a^2 r^2 - r^4 + 2 a^4 \cos (2\theta) \right]
\,,
\label{29VII22.3a.2}
\\
	\sigmaSq_{r\theta} &=
		 \sigmaSi_{r\theta} + 2 B  a^2  (a^4 - r^4) \sin (2 \theta)
\,,
\label{29VII22.3b.2}
\\
	\sigmaSq_{\theta\theta} &=
		 \sigmaSi_{\theta\theta} - B \left[ a^6 + 5 a^4 r^2 + a^2 r^4 + r^6 - 2 a^2 (a^4 + a^2 r^2 + 2 r^4) \cos (2 \theta)\right]
\,,
\label{29VII22.3c.2}
\end{align}
with $B := \tfrac{\tilde \P}{\pi} (a^2 - r^2) \left[ a^4 + r^4 - 2 a^2 r^2 \cos (2 \theta)\right]^{-2}$.
Again, visualizing the result gives a clear picture. See Figure~\ref{fig:stress_squeeze}.

\begin{figure}[t]
  \centering
  \includegraphics[scale=0.4]{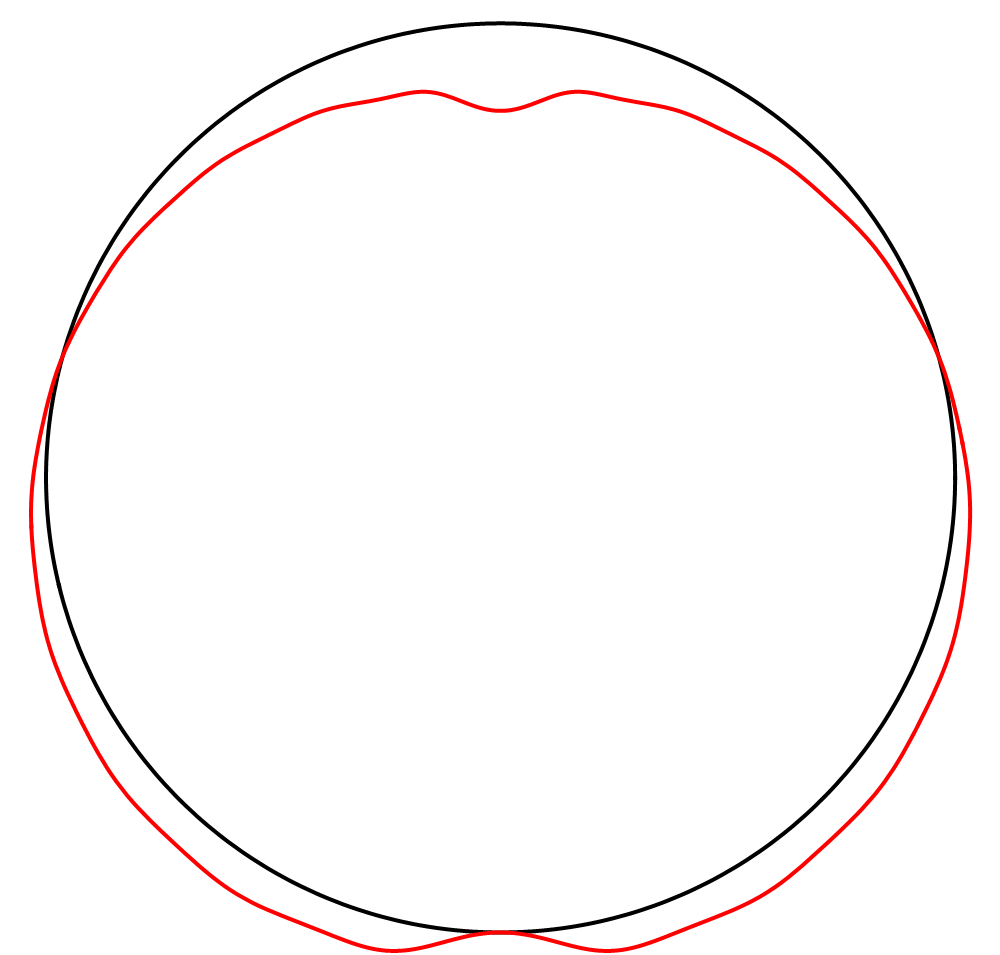}
  \caption{Illustrative deformation for the case of a waveguide squeezed between two lines, after truncating the sums in \eqref{eq:displacement_admissible_ur}-\eqref{24XI23.p1} to $n=20$, drawn in red. The undeformed reference is drawn in black.}
  \label{fig:deformation_squeeze}
\end{figure}

As before, the deformation diverges at the ``contact wires''. A truncated sum is shown in Figure~\ref{fig:deformation_squeeze}.

\subsection{Hertz contact deformations}\label{sec:Hertz}

The proper solution for the contact problem in linear elasticity is given by
the Hertz contact deformation \cite{Hertz1882}, with a modern derivation given by \cite[Chapter~9]{Landau86}.
In the special case of two cylinders contacting length-wise the two-dimensional contact region extends to a contact strip, which can be described as a contact line in terms of the cross-section (cf. \cite[Problem~2 of Chapter~9]{Landau86}).
The total force $\FHertz$
per unit length pushing two bodies into each other is distributed over a region $\{x : -s \leq x \leq s\}$, where
\begin{equation}
	s = \sqrt{\frac{4 \FHertz}{\pi} \left(\frac{1- \poisson^2}{E} + \frac{1 - \poisson'^2}{E'}\right) \frac{R R'}{R+ R'}}\,,
	\label{eq:28XI23.1}
\end{equation}
with $\{\poisson, E , R\}$ the material properties and radius of curvature of the body in question and $\{\poisson', E' , R'\}$ for the body in contact.
The pressure distribution over the contact region,   which we denote by $P_y$, is taken to be
\begin{equation}
	P_y = \frac{2 \FHertz}{\pi s} \sqrt{1- \frac{x^2}{s^2}}\,.
\end{equation}
This can be restated in the form of boundary conditions
\begin{align}
	\t{\sigma}{_r_\theta}|_{\p U} &= 0\,,
	\label{eq:Hertz_sigma_rt}
		\\
	\t{\sigma}{_r_r}|_{\p U} &=
		\left\{
			\begin{array}{ll}
				\frac{2 \FHertz}{\pi a \tan (\T)} \sqrt{1-\frac{\tan^2 (\theta)}{\tan^2 (\T)}} &  -\T \leq \theta \leq \T
					\\
				0 & \mbox{otherwise}
			\end{array}
  			\right.\,,
  			\label{eq:Hertz_sigma_rr}
\end{align}
where
\begin{equation}
	\T = \arctan \left( \frac{s}{a} \right)\,.
	\label{eq:28XI23.2}
\end{equation}

For the case of an infinite cylinder resting on a rigid plane, we simply take $R', E' \to \infty$ and $\poisson' \to 0$.

\begin{figure}[t]
  \centering
  \includegraphics[scale=0.5]{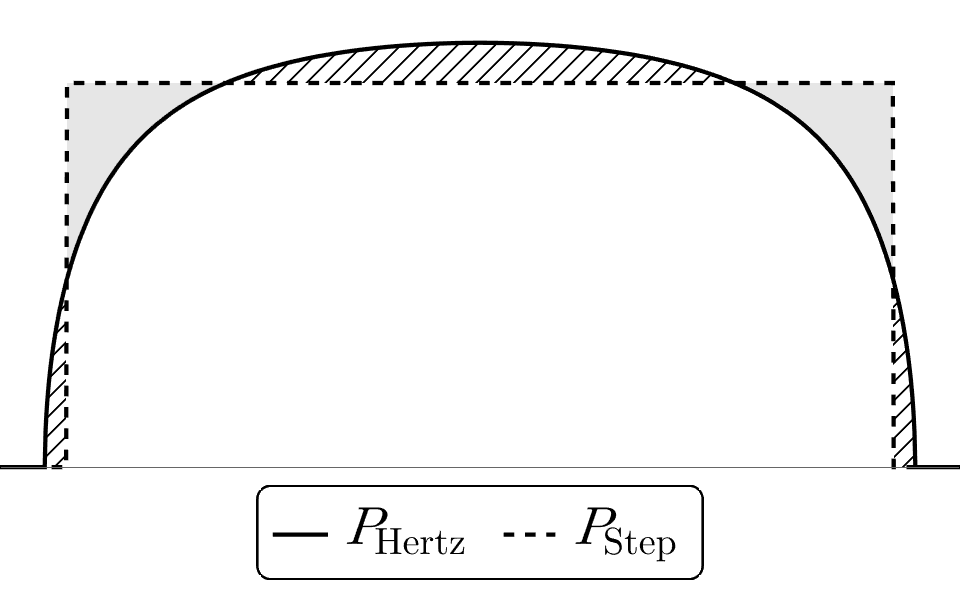}
  \caption{Comparison of the pressure distribution in the contact region for the exact Hertz solution \eqref{eq:Hertz_sigma_rr} and an approximation using  constant pressure in the contact region. The shaded and dashed regions cause overshoot and undershoot   in the  displacement figures below.}
  \label{fig:Hertz}
\end{figure}

Finding the cosine expansion of \eqref{eq:Hertz_sigma_rr} is not  obvious. Instead, one can approximate the Hertz profile with a step-function as in Figure~\ref{fig:Hertz}, with straightforward cosine expansion, allowing us to find exact expressions for the Airy functions and stresses in the settings discussed below.
The  half-width of the step function can be determined, based on the Hertz solution, to
\begin{align}
	s_\text{Hertz} &= \sqrt{\frac{4 a \FHertz}{\pi}\frac{1- \poisson^2}{E} }\,,
	\\
	\T_\text{Hertz} &= \arctan \left( \frac{s_\text{Hertz}}{a} \right)\,,
\end{align}
via \eqref{eq:28XI23.1} and \eqref{eq:28XI23.2}, ensuring a physically motivated boundary pressure distribution in response to external forces.

This approach yields exact solutions closely matching the full Hertz problem, avoiding the undesirable behaviour at the contact point exhibited by the solutions above.
We expect that this approach is sufficient for weak forces and away from the contact region.

\subsubsection{Resting on a rigid plane}\label{sec:step}

Approximating the Hertz contact solution of Section~\ref{sec:Hertz} by a step-function of constant pressure $\P $ for a single contact region from below leads to the boundary conditions
\begin{align}
	\t{\sigma}{_r_\theta}|_{\p U} &= 0\,,
		\\
	\t{\sigma}{_r_r}|_{\p U}
 &=  \P  \chi_{\left[-\T,   \T \right]} (\theta)  - \mathfrak{p}
  \big( 1- \chi_{\left[-\T,   \T \right]} (\theta) \big)
   =: \P'   \chi_{\left[-\T,   \T \right]} (\theta)  - \mathfrak{p} \,,
\end{align}
where
\begin{equation}
	\chi_A (x) := \left\{ \begin{matrix} 1 & x \in A \\ 0 & \text{else}\end{matrix}\right.
	\,,
\end{equation}
is the indicator function and $\T \approx \T_\text{Hertz}$.

A  Fourier cosine series for the boundary-values of the Airy function leads to the following coefficients in \eqref{eq:Michell_adapted}:
%
%\ptc{checked}
%
\begin{align}
	D_0 &= \frac{\P' \T}{2\pi}  - \frac{\mathfrak{p}}{2}\,,
		\\
	C_1 &= 0\,,
		\\
	A_n &= -\frac{\P' \sin (n \T)}{\pi a^{n-2} n (n-1)}\,,
		\\
	C_n &= \frac{\P' \sin (n \T)}{\pi a^n n (n+1)}\,,
\label{eq:coeffsingle}
\end{align}
with
\begin{equation}
	{\P'} = - \frac{ a \grav \rho \pi }{2 \sin (\T)}
\,.
\label{1VIII22.2.single}
\end{equation}
The summed expression for the Airy function is
\begin{align}
\notag
	\phi_\text{single} =\,&
  \frac 14 \grav \rho \Big[ (2 a^2 + r^2) r \cos (\theta) - a r^2 \frac{\T}{\sin (\T)}+  \frac{a}{\sin (\T)} ( \zeta (\theta - \T) - \zeta (\theta + \T)\Big]
\\
&- \frac 12 r^2 \mathfrak{p}
\,,
\end{align}
where the subscript ``single'' refers to a single contact region, with
\begin{equation}
	\zeta(x) := \left[ a^2 + r^2 - 2 a r \cos (x) \right] \arctan \left(\frac{r \sin (x)}{a - r \cos (x)}\right)\,.
\end{equation}

\begin{figure}[t]
  \centering
  \includegraphics[scale=0.4]{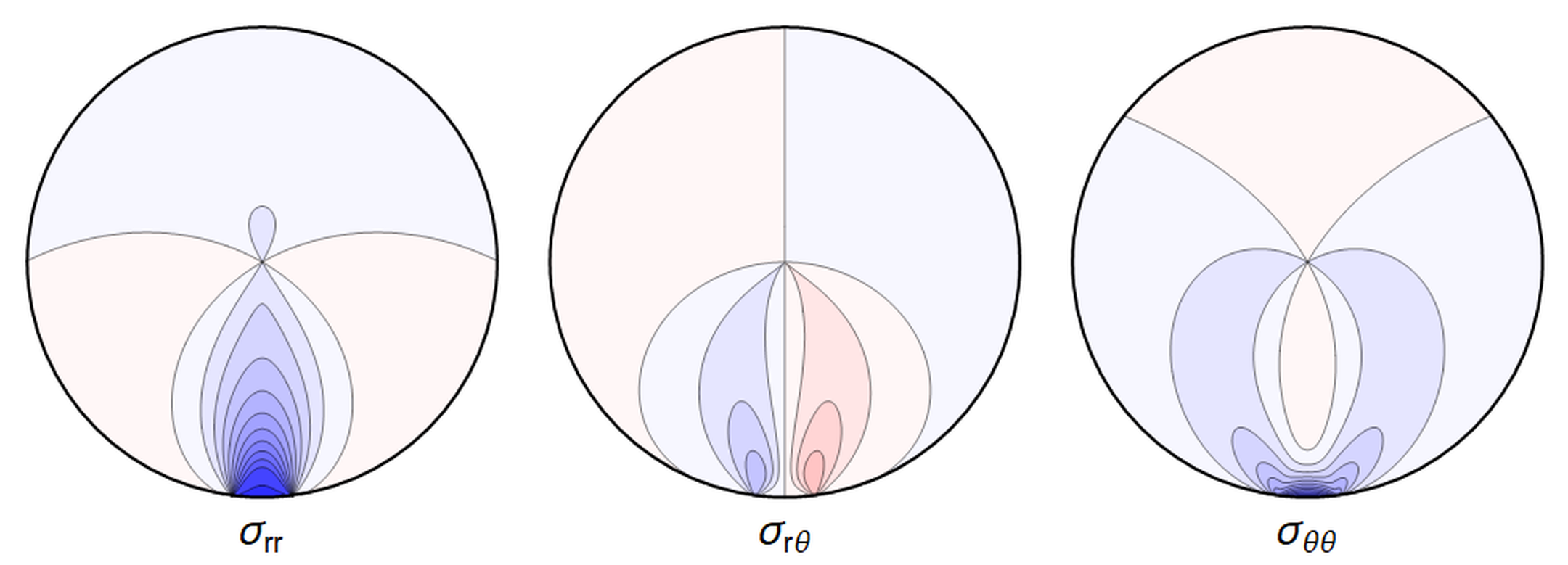}
  \includegraphics[scale=0.4]{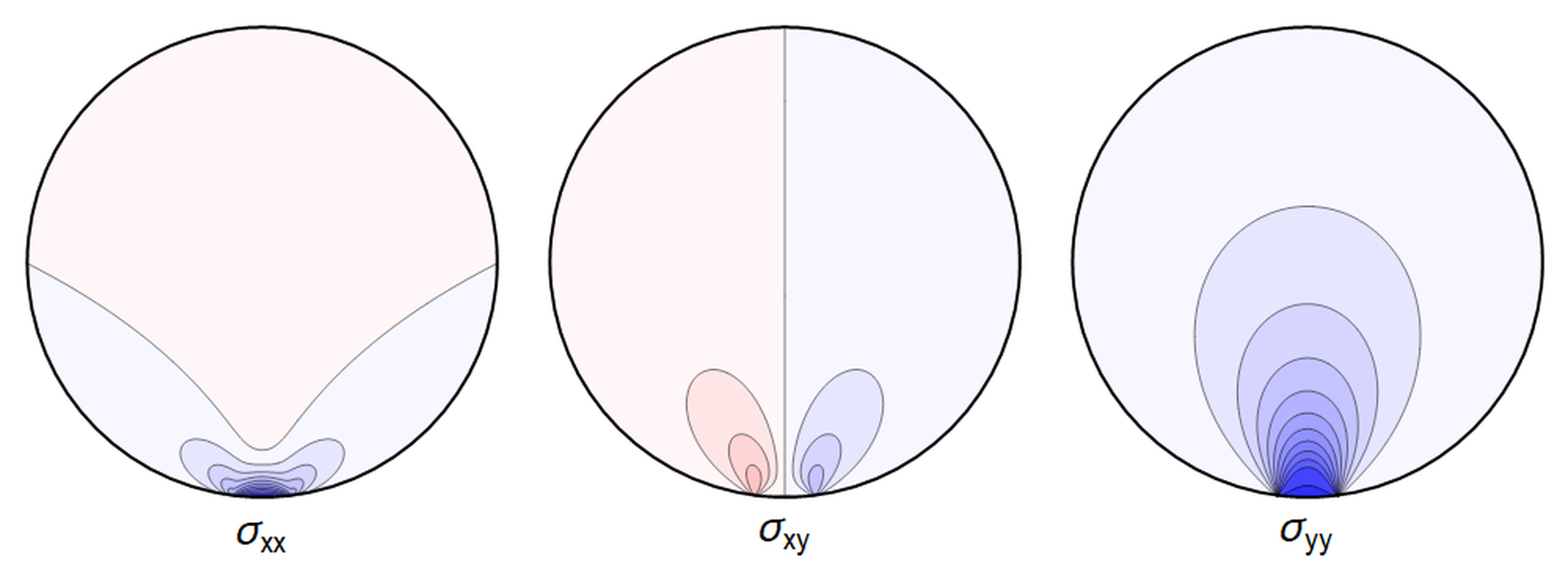}
  \caption{Typical internal stresses for a waveguide resting on an extended contact zone. Here, and in the following figures, we use the same  parameters and the same color coding as in Fig.~\ref{fig:stress_line}.}
  \label{fig:stress_plane}
\end{figure}

\begin{figure}[t]
  \begin{center}
  \includegraphics[scale=0.4]{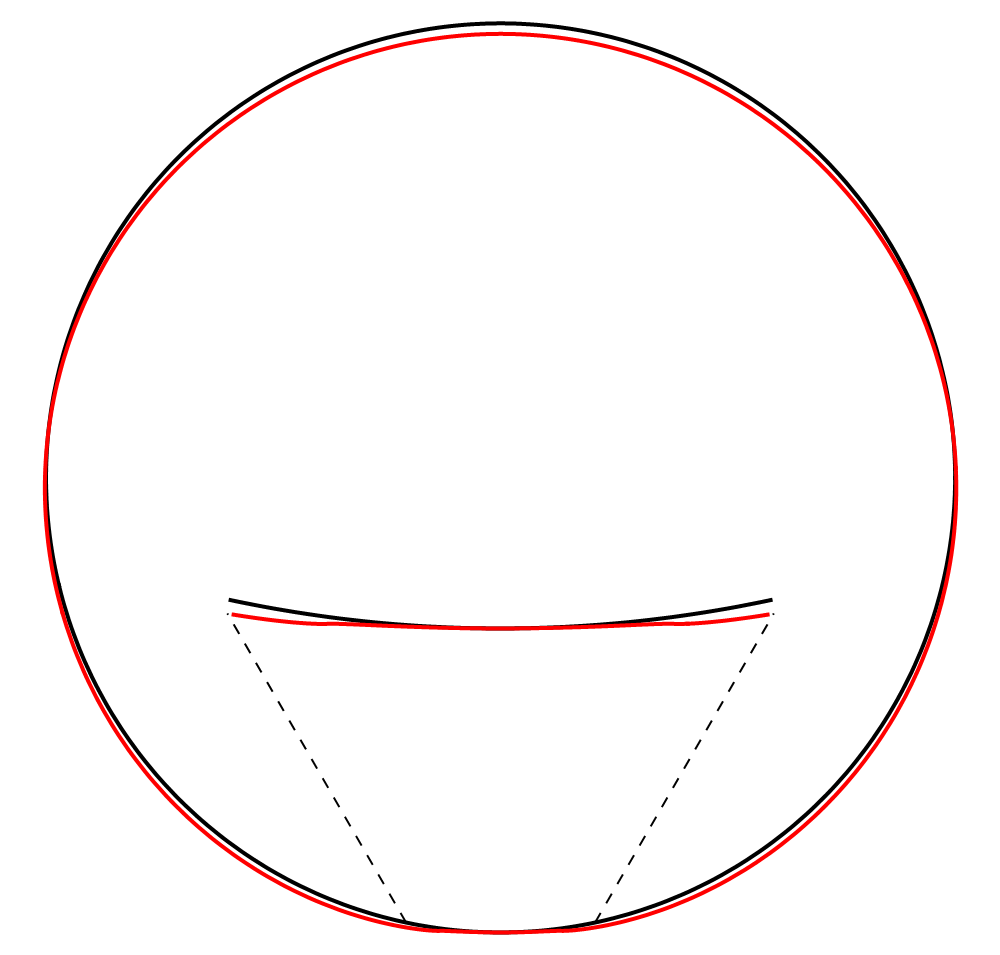}
  \end{center}
  \caption{Deformation for the case of a waveguide resting on a rigid plane, drawn in red. The undeformed reference is drawn in black. Here, and in similar figures that follow, the indentation is an artefact of the approximation illustrated in Figure~\ref{fig:Hertz}. }
  \label{fig:deformation_plane}
\end{figure}

All the sums converge, leading to an admissible displacement field.
The stresses and the displacements are shown in Figures~\ref{fig:stress_plane} and \ref{fig:deformation_plane}.


\subsubsection{Squeezed between two rigid planes}\label{sec:planesq}

The next simplest description of a spooled waveguide is one with  extended contact regions both   above and below. We model this by squeezing the fiber between two rigid planes in the spirit of Section~\ref{sec:step}, i.e. boundary conditions
\begin{align}
	\t{\sigma}{_r_\theta}|_{\p U} &= 0\,,
	\\
\notag	\sigma_{rr}|_{\p U} &= \P \chi_{\left[- \T,  \T\right]} (\theta) + \tilde \P \chi_{\left[\pi-\tilde \T, \pi + \tilde \T\right]} (\theta) - \mathfrak{p}\left[1-\chi_{\left[-\T,  \T\right]} (\theta)-\chi_{\left[\pi-\tilde \T, \pi \tilde \T\right]} (\theta) \right]
 \\
  &=: \P' \chi_{\left[-\T,  \T\right]} (\theta)
   + \tilde \P' \chi_{\left[\pi-\tilde \T, \pi
   + \tilde \T\right]} (\theta)
    - \mathfrak{p} \,,
\end{align}
with $\T$ now an angle which approximates half of the contact region from below and $\tilde \T$ the same from above.

The Fourier coefficients for the Airy function are found to be
\begin{align}
	D_0 &= \frac{\P' \T + \tilde {\P'} \tilde \T}{2\pi} - \frac{\mathfrak{p}}{2}\,,
		\\
	C_1 &= 0\,,
		\\
	A_n &= -\frac{\P' \sin (n \T) + (-1)^n \tilde {\P'} \sin (n \tilde \T)}{\pi a^{n-2} n (n-1)}\,,
		\\
	C_n &= \frac{\P' \sin (n \T) + (-1)^n \tilde {\P'} \sin (n \tilde \T)}{\pi a^n n (n+1)}\,,
\label{eq:coeffdouble}
\end{align}
with
\begin{equation}
	{2 \sin (\T)}{\P'}
 =
 2 \tilde{\P'} \sin (\tilde \T) -  { a \grav \rho \pi }
\,.
\label{1VIII22.2.double}
\end{equation}

The re-summed expression for the Airy function is
\begin{align}\notag	
	\phi_\text{double} = \phi_\text{single}
	+ \frac{\tilde {\P'}}{2 \pi} \Bigg[&r^2 \left(\tilde \T + \T \frac{\sin (\tilde \T)}{\sin (\T)}\right)
	+ \Big(\xi(\theta - \tilde \T) - \xi (\theta +\tilde \T) \Big)
	\\
&- \frac{\sin (\tilde \T)}{\sin (\T)} \Big(\zeta (\theta - \T)  - \zeta (\theta + \T)\Big)
	 \Bigg]\,,
\end{align}
where
\begin{equation}
	\xi(x) := \left[ a^2 + r^2 + 2 a r \cos (x) \right] \arctan \left(\frac{r \sin (x)}{a + r \cos (x)}\right)\,.
\end{equation}

\begin{figure}[t]
  \centering
  \includegraphics[scale=0.4]{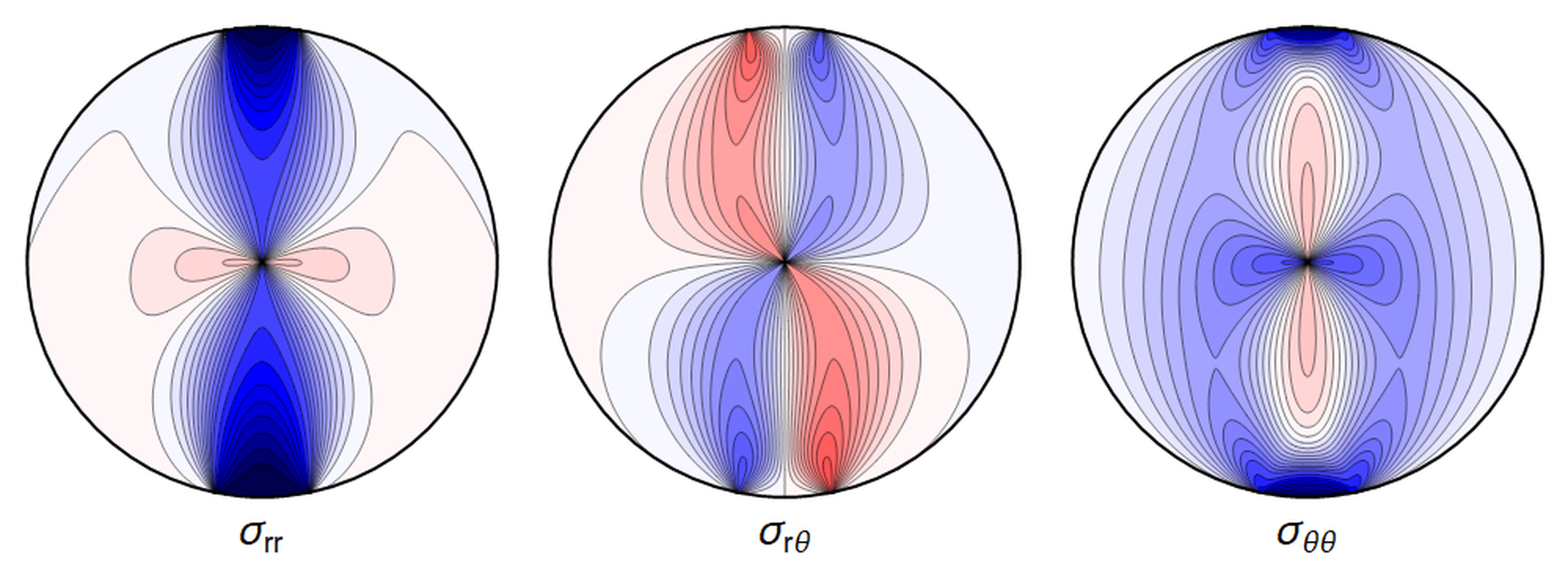}
  \includegraphics[scale=0.4]{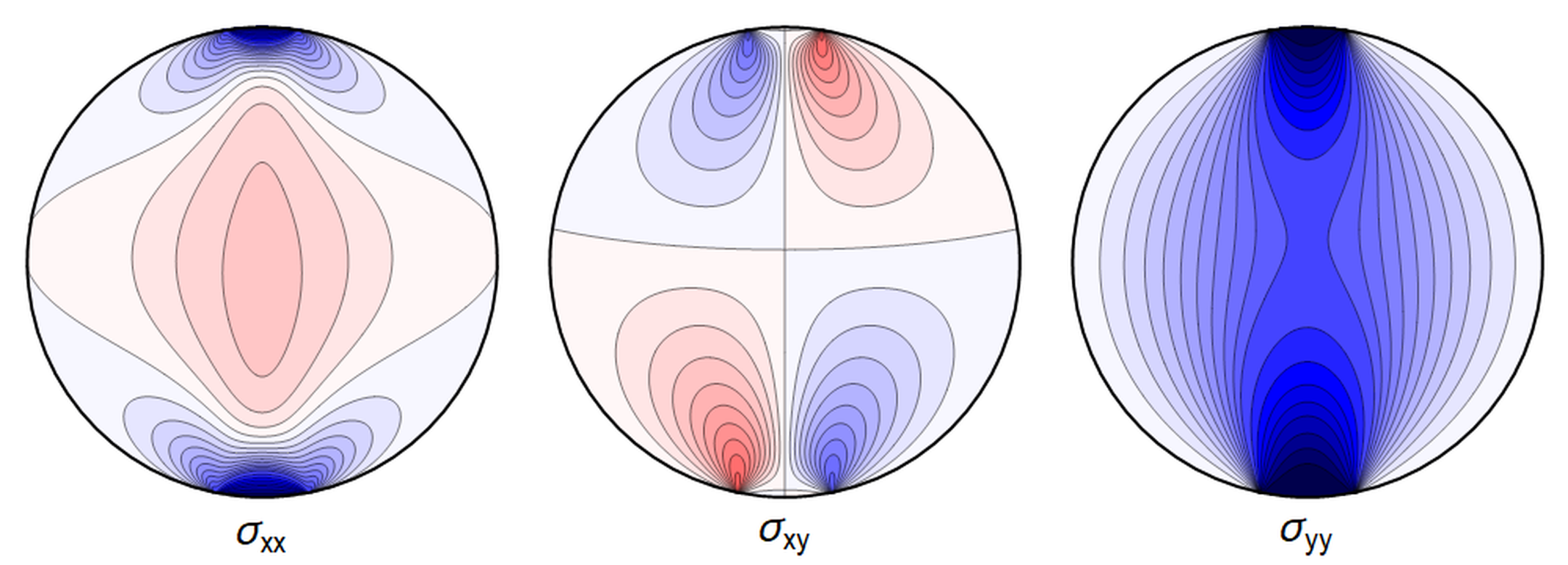}
  \caption{Internal stresses for the case of a waveguide squeezed between two rigid planes with extended contact region.}
  \label{fig:stress_plane_double}
\end{figure}

\begin{figure}[t]
  \centering
  \includegraphics[scale=0.4]{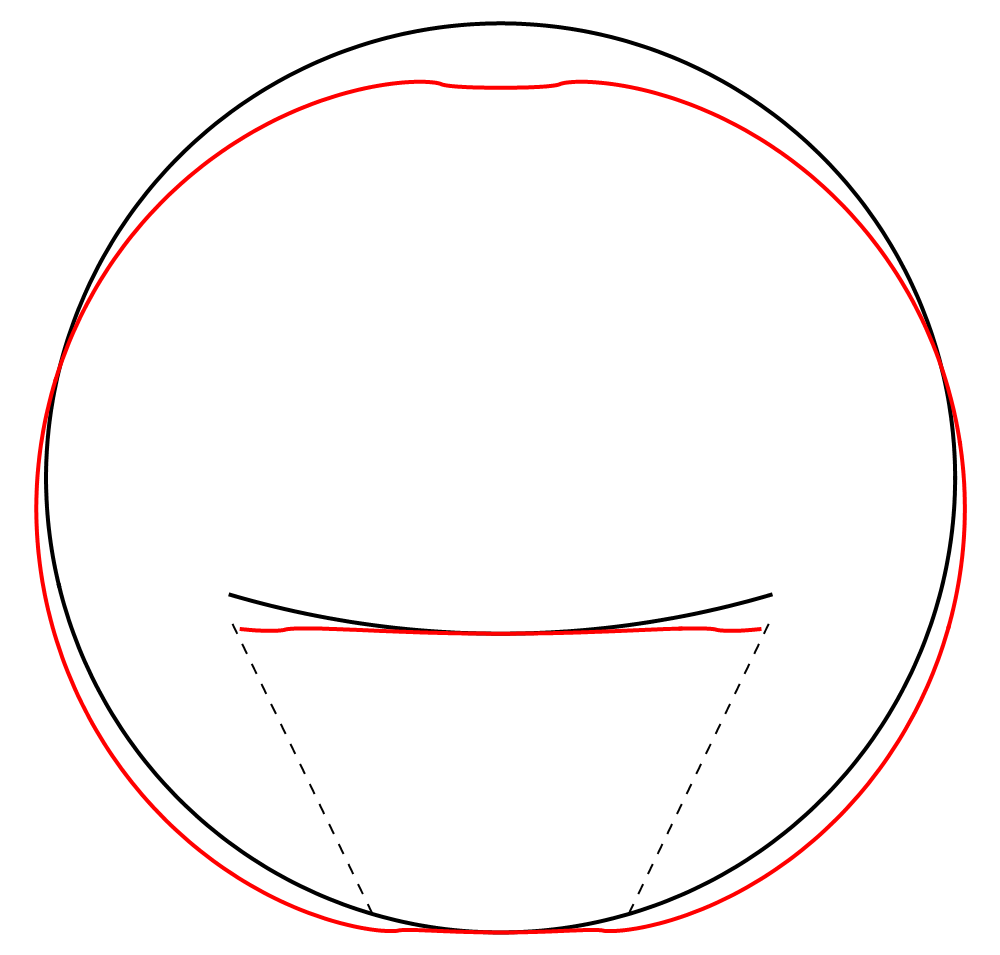}
  \caption{Deformation for the case of a waveguide squeezed between rigid planes of finite width, drawn in red. The undeformed reference is drawn in black. }
  \label{fig:deformation_plane_double}
\end{figure}

All the sums converge, leading to a finite displacement field.
The stresses and the displacements can be seen in Figure~\ref{fig:stress_plane_double} and Figure~\ref{fig:deformation_plane_double} respectively.
 Here, and in the plots that follow, the parameters  are related to those of the two contact-lines case via the correspondence
\begin{equation}
 \tilde \P_{\text{contact line}} =   \tilde \P_{\text{extended contact region}} \times 2  \tilde \Theta
 \,,
\label{20I24.1}
\end{equation}
which guarantees an identical integrated reaction force from the support.
\begin{figure}[htb]
  \centering
  \includegraphics[scale=0.4]{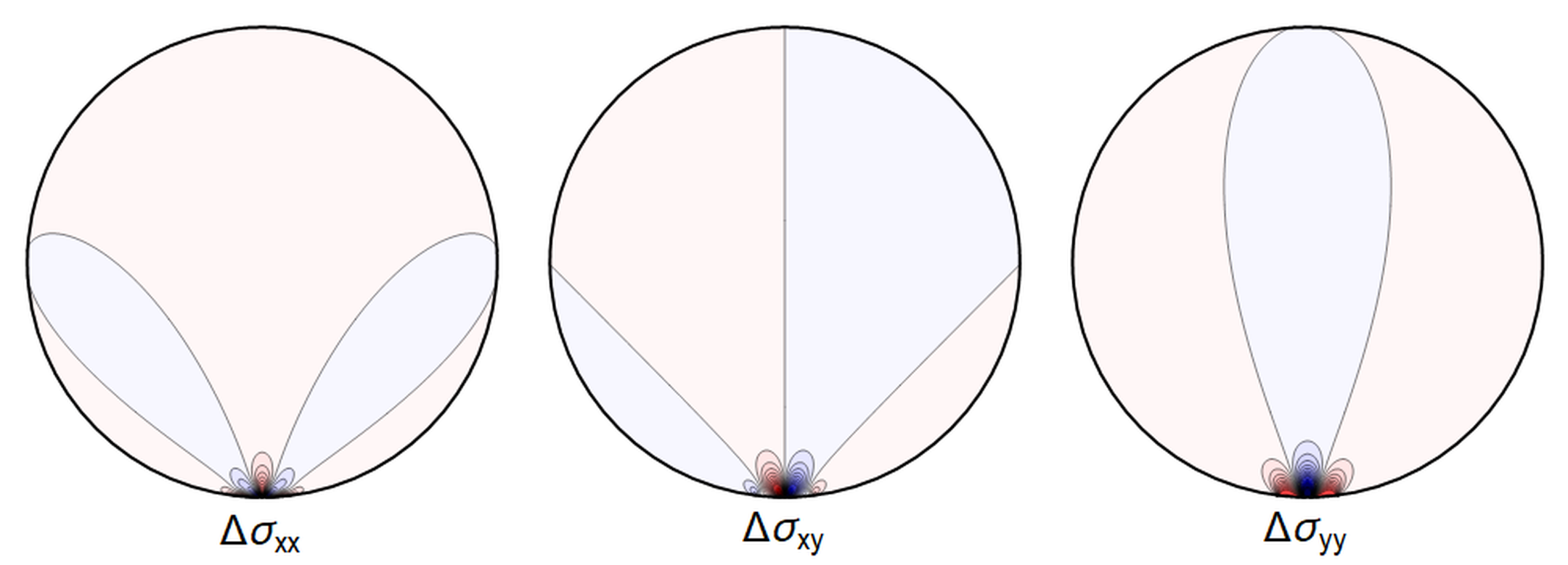}
  \caption{The difference $\Delta \sigma$ between the line contact Fig.~\ref{fig:stress_line} and extended region Fig.~\ref{fig:stress_plane} cases.}
  \label{fig:stress_compare_single}
\end{figure}
\begin{figure}[htb]
  \centering
  \includegraphics[scale=0.4]{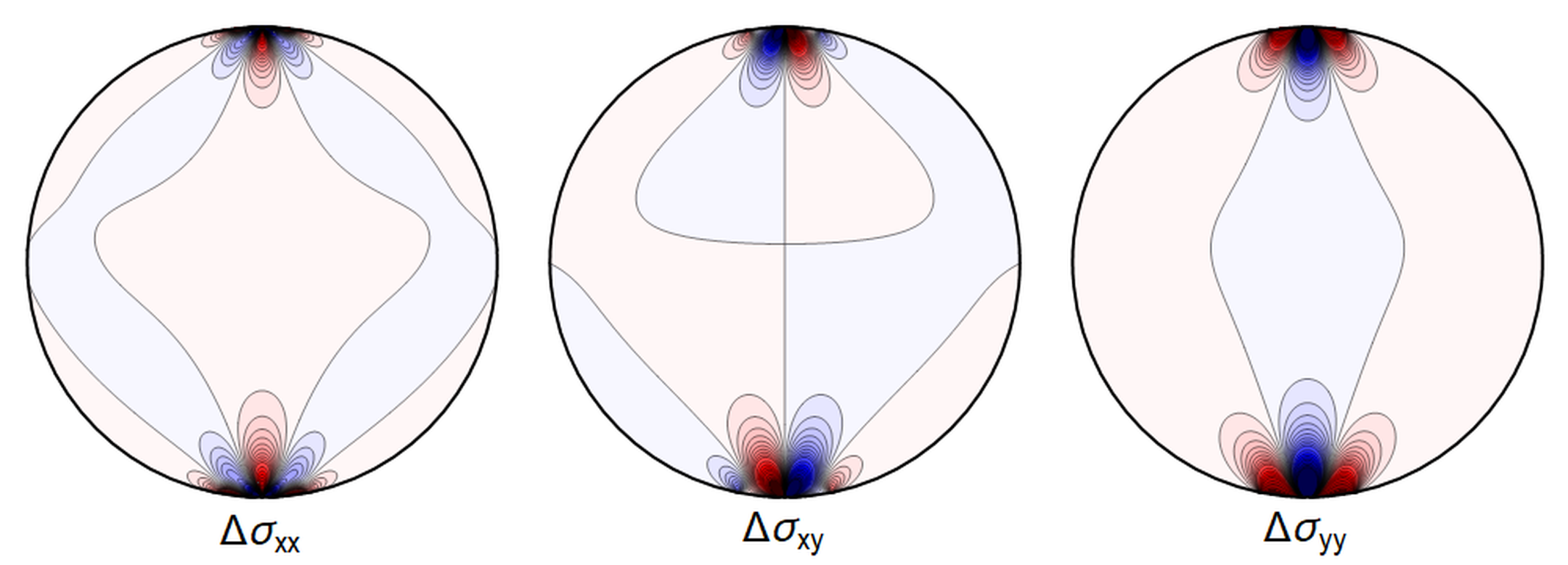}
  \caption{Difference $\Delta \sigma$ between the line contact of Fig.~\ref{fig:stress_squeeze} and the extended region of Fig.~\ref{fig:stress_plane_double}.}
  \label{fig:stress_compare_double}
\end{figure}
It is of interest to compare the line-contact cases to the corresponding extended region cases, focusing on the central region containing the guiding core of the waveguide. In Figures~\ref{fig:stress_compare_single} and \ref{fig:stress_compare_double}
we show the differences for the same set of parameters as used in the plots.
We can quantify the differences away from the contact points by comparing the maximal values over the regions
$$U_{1/2}= \{(r,\theta) : 0 < r\leq \tfrac{a}{2}, - \pi < \theta \leq \pi \}
$$
and
$$U_{1/10}= \{(r,\theta) : 0 < r\leq \tfrac{a}{10}, - \pi < \theta \leq \pi \}
 \,,
$$
%,
and setting
\begin{equation}
	\Delta_U (\sigma_1, \sigma_2)= \frac{2 \max_{U} |\sigma_1 - \sigma_2|}{\max_{U} |\sigma_1| + \max_{U} |\sigma_2|}\,.
\end{equation}
The results, for the  parameters of Table~\ref{tab:visual_parameters} (used in all our figures), are shown in Tables~\ref{tab:stress_compare_single_alt} and \ref{tab:stress_compare_double_alt}.
\begin{table}[h!]
$$
\begin{array}{||c||c|c|c|c|c|c||}
\hline\hline
		&	\sigma_{rr}	&	\sigma_{r\theta} 	&	\sigma_{\theta\theta} & \sigma_{xx} & \sigma_{xy} & \sigma_{yy}	
	 \\\hline
	 \Delta_{U_{1/2}}	& 	2.9\%	& 	4.4\%	 &	5.5\% &	11.7\%	&	6.5\%	&	2.9\%
	 \\\hline
	  \Delta_{U_{1/10}}	&	0.6\%	 &  	1.1\%	 & 	0.9\%	&	2.5\%	&	2.3\%	&	0.6\%
	 \\ \hline\hline
\end{array}
$$
\caption{Relative differences for the single contact region models.}
\label{tab:stress_compare_single_alt}
\end{table}
\begin{table}[h!]
$$
\begin{array}{||c||c|c|c|c|c|c||}
\hline\hline
		&	\sigma_{rr}	&	\sigma_{r\theta} 	&	\sigma_{\theta\theta}	& \sigma_{xx} & \sigma_{xy} & \sigma_{yy}	
	 \\\hline
	 \Delta_{U_{1/2}}	& 	2.4\%	& 	9.9\%	 &	7.7\%	&	11.0\%	&	17.2\%	&	3.9\%
	 \\\hline
	  \Delta_{U_{1/10}}	&	1.4\%	 &  	2.0\%	 & 	1.7\%	&	5.1\%	&	10.4\%	&	1.5\%
	  \\\hline\hline
\end{array}
$$
\caption{Relative differences for the double-contact-region models.}
\label{tab:stress_compare_double_alt}
\end{table}

A reasonable approximation is obtained in the central region for the values of parameters considered if a 11 \% error is less than the measurement errors at hand.

\subsubsection{Four contact regions}
\label{sec: quadrilateral}

We consider now a configuration as in Figure~\ref{fig:coilquad},
with four contact regions for the waveguide. Using step-function boundary-stresses for the contact region gives
\begin{align}
	\t{\sigma}{_r_\theta}
|_{\p U} =\,&  0\,,
\end{align}
\begin{align}
\notag	\sigma_{rr} |_{\p U} =\,&\P \chi_{\left[-\T,   \T\right]} (\theta) + \tilde \P \chi_{\left[\pi-\tilde \T, \pi + \tilde \T\right]} (\theta) +\bar \P  \Big[\chi_{\left[\tfrac{\pi}{2}-\hat \T, \tfrac{\pi}{2} + \hat \T\right]} (\theta) + \chi_{\left[\tfrac{3\pi}{2}-\hat \T, \tfrac{3 \pi}{2} + \hat \T\right]} (\theta) \Big]
		\\\notag
&- \mathfrak{p}\Big[1 - \chi_{\left[-\T,   \T\right]} (\theta) -  \chi_{\left[\pi-\tilde \T, \pi + \tilde \T\right]} (\theta)- \chi_{\left[\tfrac{\pi}{2}-\hat \T, \tfrac{\pi}{2} + \hat \T\right]} (\theta) - \chi_{\left[\tfrac{3\pi}{2}-\hat \T, \tfrac{3 \pi}{2} + \hat \T\right]} (\theta) \Big]
\\
\notag =:\,& \P' \chi_{\left[- \T,   \T\right]} (\theta) + \tilde \P' \chi_{\left[\pi-\tilde \T, \pi + \tilde \T\right]} (\theta) +{\bar \P}'  \Big[\chi_{\left[\tfrac{\pi}{2}-\hat \T, \tfrac{\pi}{2} + \hat \T\right]} (\theta) + \chi_{\left[\tfrac{3\pi}{2}-\hat \T, \tfrac{3 \pi}{2} + \hat \T\right]} (\theta) \Big]
\\
&- \mathfrak p
\,,
\end{align}
with $\T$ now an angle which approximates half of the contact region from below and $\tilde \T$ the same from above and $\hat \T$ from the sides.
\begin{figure}[t]
  \centering
  \includegraphics[scale=0.4]{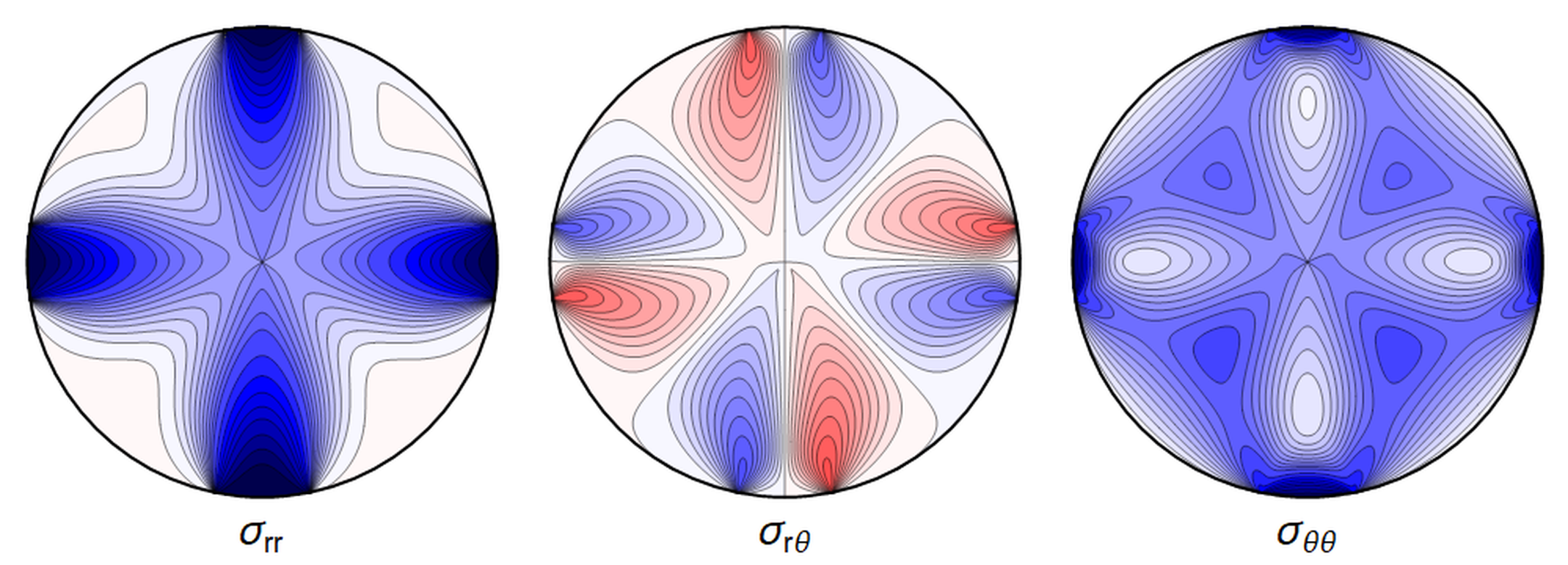}
  \includegraphics[scale=0.4]{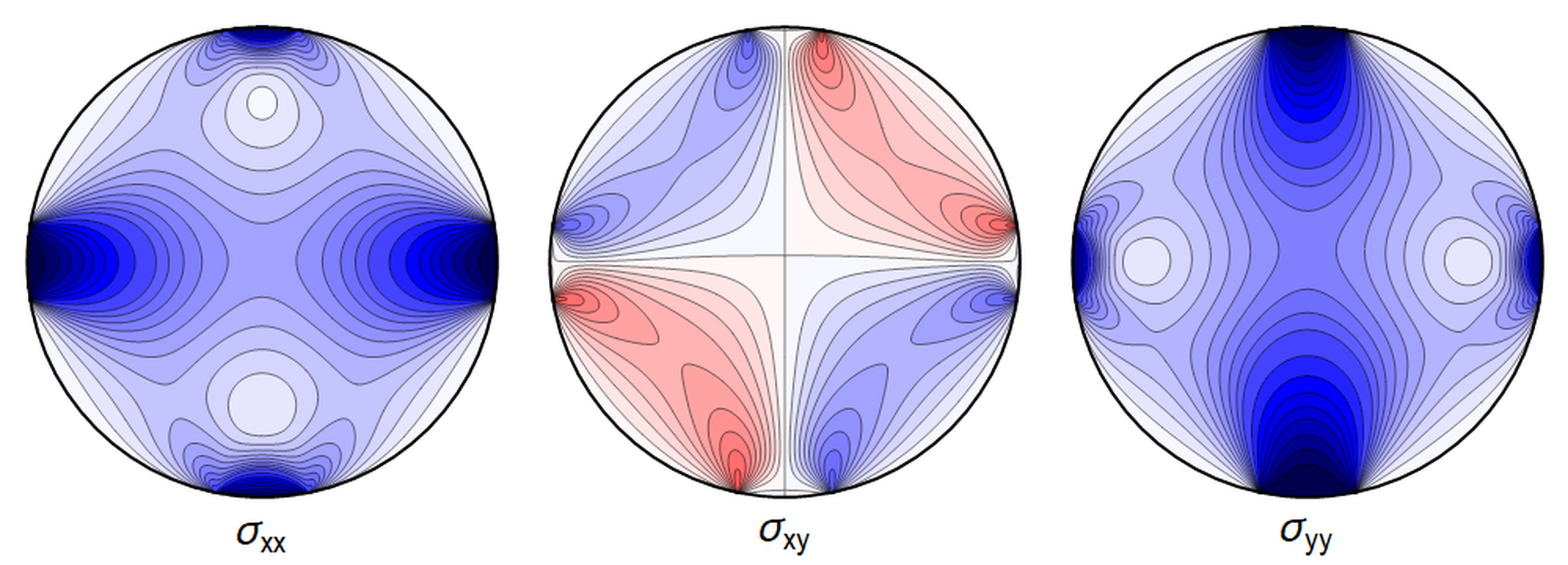}
  \caption{Internal stresses for the case of a waveguide squeezed between four rigid planes  from below, above and the sides as depicted in Figure~\ref{fig:coilquad}, with extended contact region. }
  \label{fig:stress_plane_quad}
\end{figure}
\begin{figure}[t]
  \centering
  \includegraphics[scale=0.4]{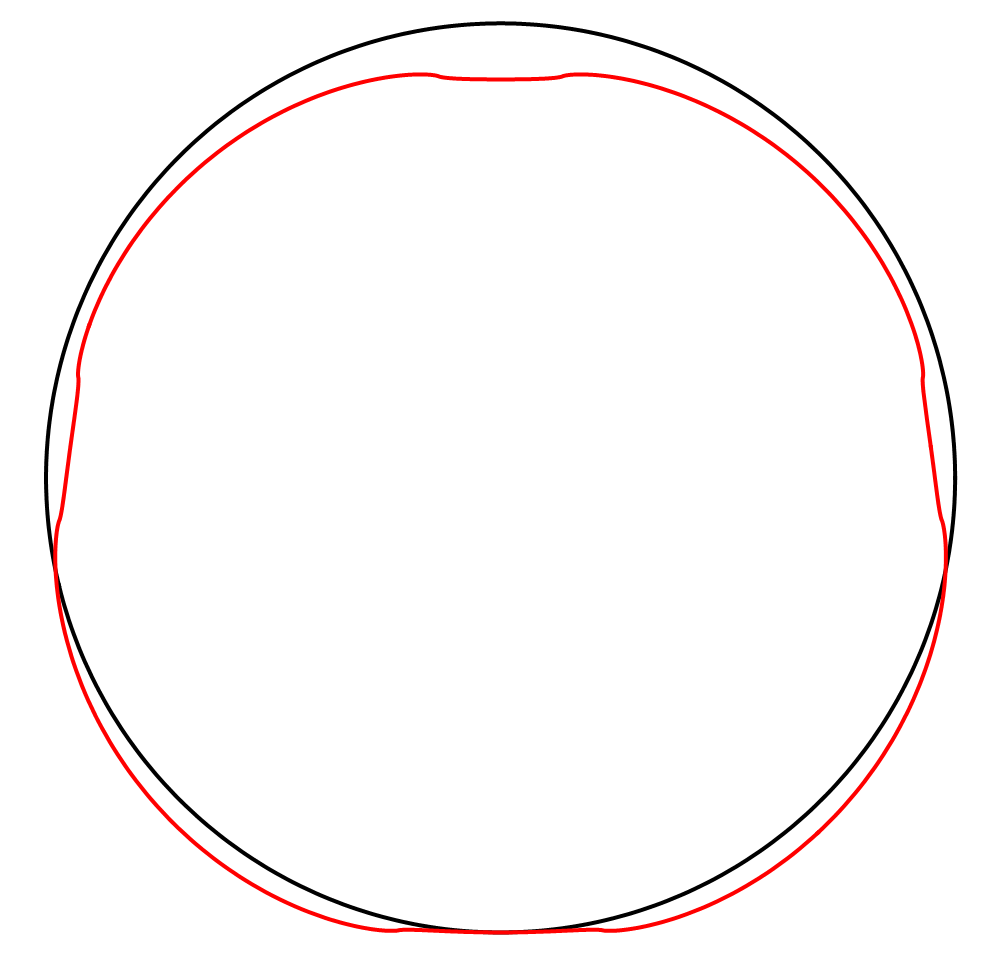}
  \caption{Deformation for the case of a waveguide squeezed between rigid planes of finite width as in Figure~\ref{fig:coilquad}, drawn in red. The undeformed reference is drawn in black. }
  \label{fig:deformation_plane_quad}
\end{figure}

The resulting Fourier coefficients in the Airy function are
\begin{align}
	D_0 &= \frac{\P' \T + \tilde \P' \tilde \T + 2 {\bar \P}' \hat \T}{2\pi} - \frac{\mathfrak{p}}{2}\,,
		\\
	C_1 &= 0\,,
		\\
	A_n &= -\frac{\P' \sin (n \T) + (-1)^n \tilde \P' \sin (n \tilde \T) + 2 {\bar \P}' \cos\left(\tfrac{n \pi}{2}\right) \sin (n \hat \T)}{\pi a^{n-2} n (n-1)}\,,
		\\
	C_n &= \frac{\P' \sin (n \T) + (-1)^n \tilde \P' \sin (n \tilde \T) + 2 {\bar \P}' \cos\left(\tfrac{n \pi}{2}\right) \sin (n \hat \T)}{\pi a^n n (n+1)}\,,
\label{eq:coeffquad}
\end{align}
with
\begin{equation}
	\P' = \tilde \P' \frac{\sin (\tilde \T)}{\sin (\T)}- \frac{ a \grav \rho \pi }{2 \sin (\T)}
\,.
\label{1VIII22.2.quad}
\end{equation}

All the sums converge, leading to an admissible displacement field.
The stresses and the displacement can again be produced graphically -- see Figure~\ref{fig:stress_plane_quad} and Figure~\ref{fig:deformation_plane_quad}.

\subsubsection{Squeezed between six rigid planes}
\label{sec: hexagonal}

Our final model for spooled waveguides is given by Figure~\ref{fig:coilhexa}, which we model by imposing the boundary conditions
\begin{align}
	\t{\sigma}{_r_\theta}|_{\p U} =\,& 0\,,
		\\
\notag	\sigma_{rr}|_{\p U} =\,& \P \chi_{\left[-\T,  \T\right]} (\theta) + \tilde \P \chi_{\left[\pi-\tilde \T, \pi + \tilde \T\right]} (\theta) +\hat \P  \Big[\chi_{\left[\tfrac{\pi}{3}-\hat \T, \tfrac{\pi}{3} + \hat \T\right]} (\theta) + \chi_{\left[\tfrac{5\pi}{3}-\hat \T, \tfrac{5 \pi}{3} + \hat \T\right]} (\theta) \Big]
	\\
\notag	&+\check \P  \Big[\chi_{\big[\tfrac{2\pi}{3}-\check \T, \tfrac{2\pi}{3} + \check \T\big]} (\theta) + \chi_{\big[\tfrac{4\pi}{3}-\check \T, \tfrac{4 \pi}{3} + \check \T\big]} (\theta) \Big]
\\
\notag &- \mathfrak{p} \Big[1 - \chi_{\left[- \T,  \T\right]} (\theta) - \chi_{\left[\pi-\tilde \T, \pi + \tilde \T\right]} (\theta) - \chi_{\left[\tfrac{\pi}{3}-\hat \T, \tfrac{\pi}{3} + \hat \T\right]} (\theta) - \chi_{\left[\tfrac{5\pi}{3}-\hat \T, \tfrac{5 \pi}{3} + \hat \T\right]} (\theta)
\\
\notag & - \chi_{\big[\tfrac{2\pi}{3}-\check \T, \tfrac{2\pi}{3} + \check \T\big]} (\theta) - \chi_{\big[\tfrac{4\pi}{3}-\check \T, \tfrac{4 \pi}{3} + \check \T\big]} (\theta)\Big]
	\\
\notag	=\,& \P' \chi_{\left[- \T,  \T\right]} (\theta) + \tilde \P' \chi_{\left[\pi-\tilde \T, \pi + \tilde \T\right]} (\theta) +\hat \P'  \Big[\chi_{\left[\tfrac{\pi}{3}-\hat \T, \tfrac{\pi}{3} + \hat \T\right]} (\theta) + \chi_{\left[\tfrac{5\pi}{3}-\hat \T, \tfrac{5 \pi}{3} + \hat \T\right]} (\theta) \Big]
	\\
	&+\check \P'  \Big[\chi_{\big[\tfrac{2\pi}{3}-\check \T, \tfrac{2\pi}{3} + \check \T\big]} (\theta) + \chi_{\big[\tfrac{4\pi}{3}-\check \T, \tfrac{4 \pi}{3} + \check \T\big]} (\theta) \Big] - \mathfrak{p}
	\,,
\label{eq: BC hexagonal}
\end{align}
with $\T$ now an angle which approximates half of the contact region from below, and $\tilde \T$ the same from above, and $\hat \T$ and $\check \T$ from the sides.

The associated Fourier cosine series for the Airy functions has   coefficients
\begin{align}
	D_0 =& \frac{\P' \T + \tilde \P' \tilde \T + 2\hat \P' \hat \T + 2 \check \P' \check \T}{2\pi} - \frac{\mathfrak{p}}{2}\,,
		\\
	C_1 =& 0\,,
		\\\notag
	A_n =& -\frac{1}{\pi a^{n-2} n (n-1)}\Big[\P' \sin (n \T) + (-1)^n \tilde \P' \sin (n \tilde \T)
		\\
		&\qquad\qquad\qquad\qquad + 2 \hat \P' \cos\left(\tfrac{n \pi}{3}\right) \sin (n \hat \T) +2 \check \P' \cos\left(\tfrac{2n \pi}{3}\right) \sin (n \check \T) \Big]\,,
		\\\notag
	C_n =& \frac{1}{\pi a^n n (n+1)} \Big[\P' \sin (n \T) + (-1)^n \tilde \P' \sin (n \tilde \T)
		\\
		&\qquad\qquad\qquad + 2 \hat \P' \cos\left(\tfrac{n \pi}{3}\right) \sin (n \hat \T) +2 \check \P' \cos\left(\tfrac{2n \pi}{3}\right) \sin (n \check \T) \Big]\,,
\label{eq:coeffhexa}
\end{align}
with
\begin{equation}
	\P' = \tilde \P' \frac{\sin (\tilde \T)}{\sin (\T)} -\hat \P' \frac{\sin (\hat \T)}{\sin (\T)} +\check \P' \frac{\sin (\check \T)}{\sin (\T)}- \frac{ a \grav \rho \pi }{2 \sin (\T)}
\,.
\label{1VIII22.2.hexa}
\end{equation}

\begin{figure}[t]
  \centering
  \includegraphics[scale=0.4]{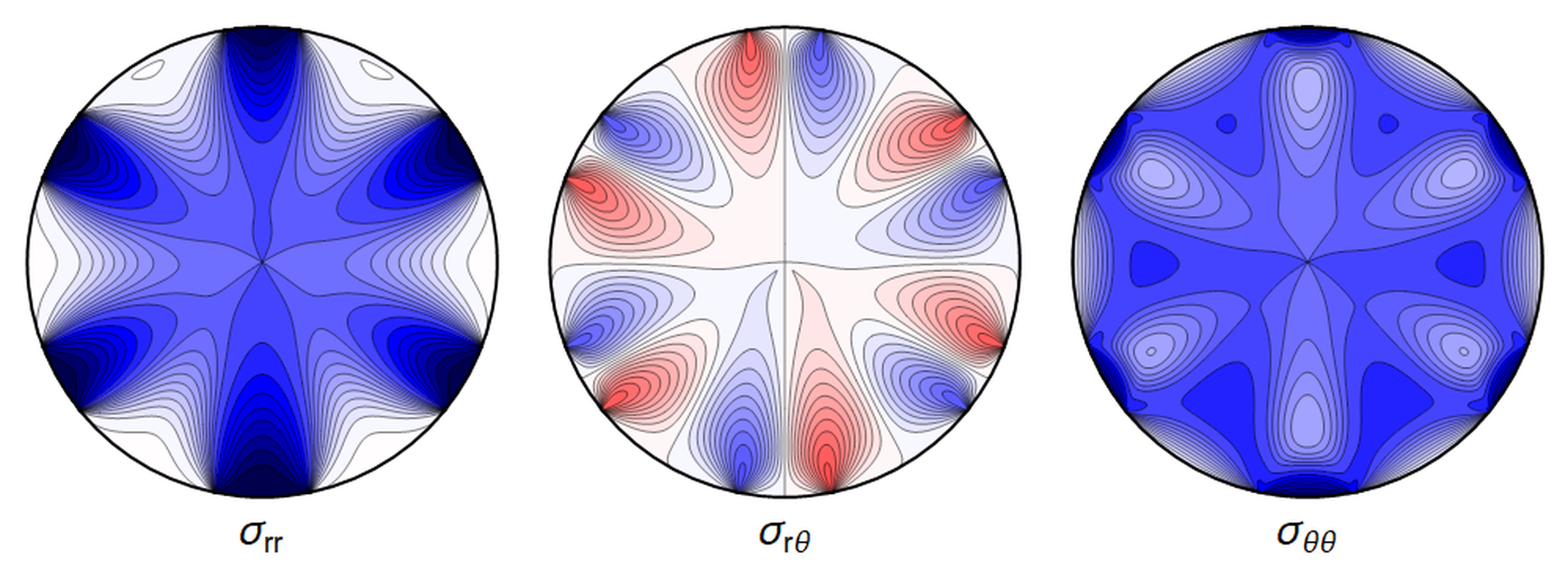}
  \includegraphics[scale=0.4]{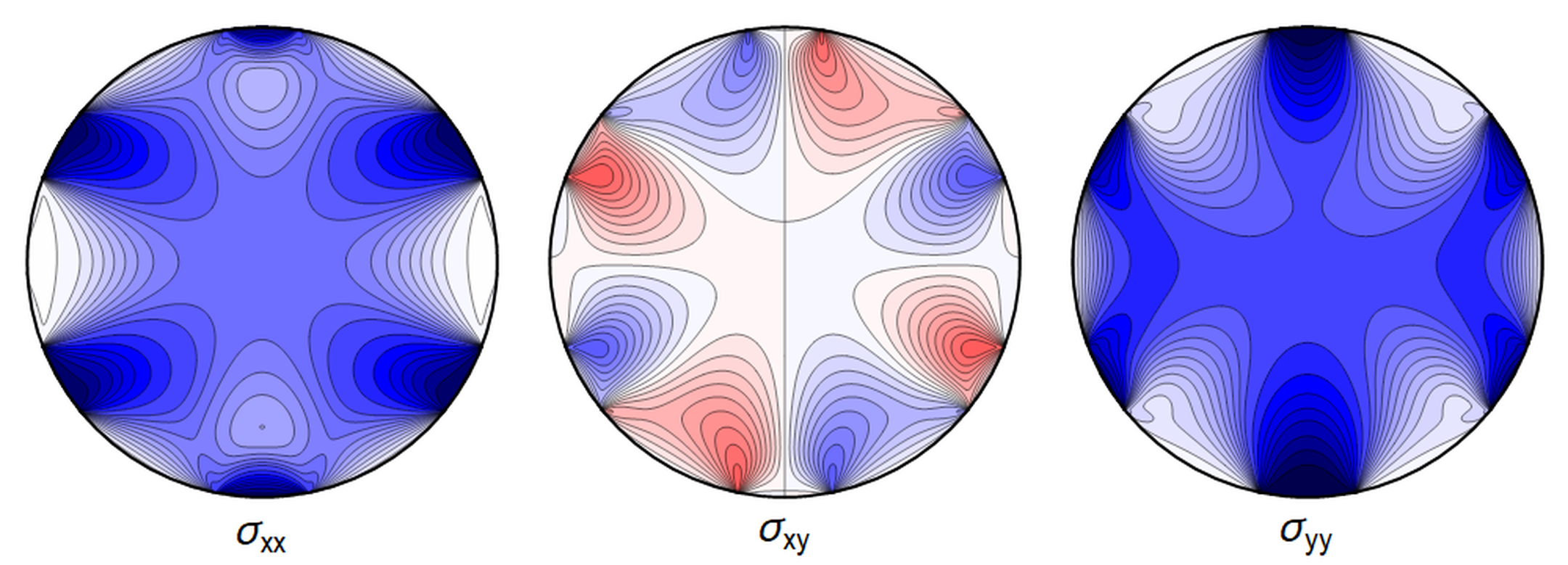}
  \caption{Internal stresses for the case of a waveguide squeezed between six
  rigid planes with extended contact regions, as in   Figure~\ref{fig:coilhexa}. }
  \label{fig:stress_plane_hexa}
\end{figure}

\begin{figure}[t]
  \centering
  \includegraphics[scale=0.4]{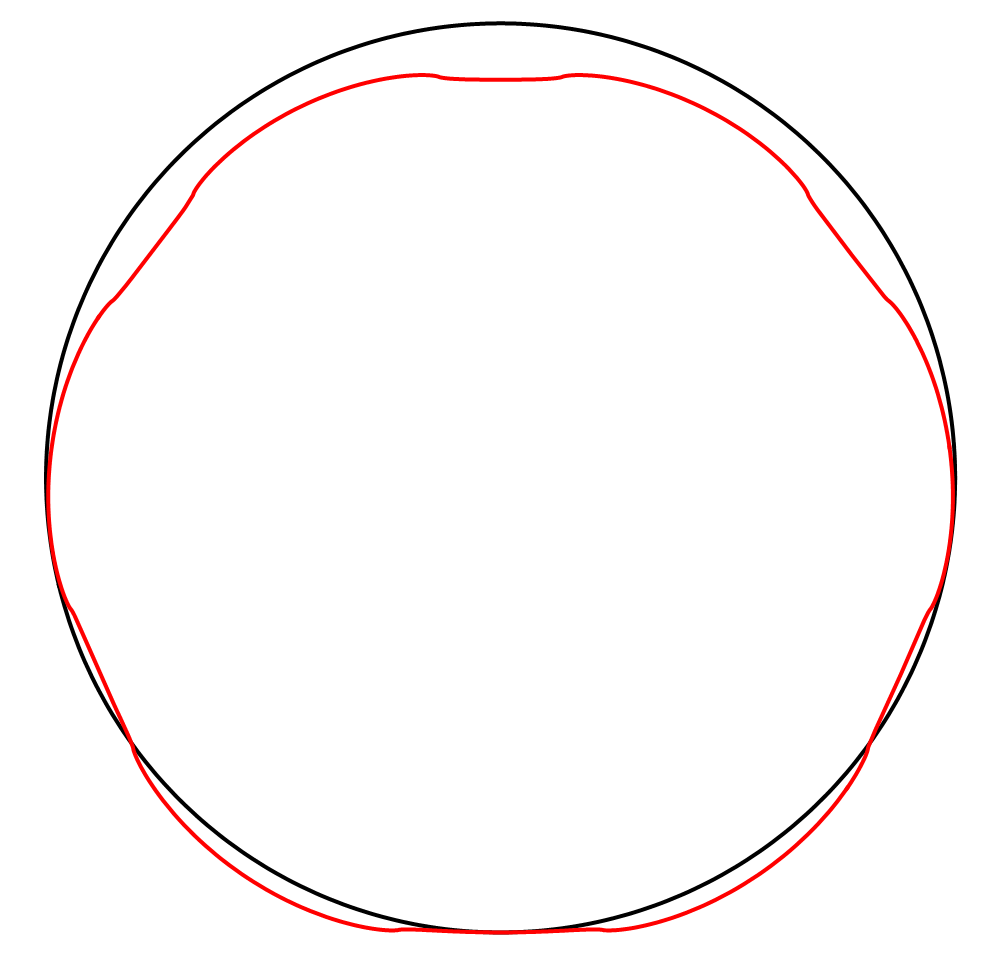}
  \caption{Deformation for the case of a waveguide squeezed between rigid planes of finite width as in Figure~\ref{fig:coilhexa}, drawn in red. The undeformed reference is drawn in black. }
  \label{fig:deformation_plane_hexa}
\end{figure}

All the sums converge, leading to a finite displacement field.
The stresses and the displacement are seen in Figures~\ref{fig:stress_plane_hexa} and \ref{fig:deformation_plane_hexa}.
\section{Estimates for GRAVITES}
\label{sec:gravites}

The planned GRAVITES experiment \cite{HMMMCW} will use single-mode optical fibers made of  silica glass, which we assume to be homogeneous, with material properties and experimental parameters given in Table~\ref{tab:SiO2}--\ref{tab:parGRAVITES}.
The numbers used below, such as the change of pressure $\delta \mathfrak p$ and the change of temperature $T-T_0$, and their outcome for the experiment,  correspond to a change in height of $1\,\mathrm{m}$ at sea level
 for an unshielded experiment.

 It should be kept in mind that the real experiment will be in a vacuum chamber. There are various outcomes possible, depending upon the details of the experimental implementation of the problem:

 \begin{enumerate}
 \item The effects scale in an obvious way with the residual pressure, in which case our calculations provide bounds on the quality of the vacuum needed for measurability of the gravitational effect.
     \item The gradients of the temperature of the residual gas affect the temperature of the waveguide during the timescale of the experiment,  in which case our calculations provide bounds on the residual temperature gradients.
\end{enumerate}

The above can of course be circumvented by ensuring that there are neither pressure gradients nor temperature gradients, of the kind considered here, inside the vacuum chamber at the timescale of the experiment, so that  the effects determined here become irrelevant. Our calculations thus give  guidance concerning the experimental setup.

\begin{table}[h!]
$$
\begin{array}{||c|c|c|c||}
\hline
\hline
	\nu	&	E	&	\rho &	 \alphat	
	 \\
\hline
	0.17	&	73.1\,\mathrm{GPa} &	2.2\,\mathrm{g/cm^3}	&	1.8\cdot10^{-7}\,\mathrm{K^{-1}}
\\
\hline
\hline
\end{array}
$$
\caption{Properties of silica oxide glass from \cite{Crystran12}.}
\label{tab:SiO2}
\end{table}
\begin{table}[h!]
$$
\begin{array}{||c|c|c|c|c|c||}
\hline
\hline
	a 	& \delta \grav	& \delta \mathfrak{p} & T-T_0 & L & \beta
	 \\
\hline
	62.5\,\mathrm{\upmu m}	&	3\,\mathrm{\upmu m/s^2}	 & 10\,\mathrm{Pa} & 10^{-2}\,\mathrm{K} & 10^5\,\mathrm{m} & 6\cdot 10^6\,\mathrm{m^{-1}}
\\
\hline
\hline
\end{array}
$$
\caption{Parameters for GRAVITES from \cite{HMMMCW}.}
\label{tab:parGRAVITES}
\end{table}

In what follows we will also need to estimate the spooling force. We choose
$$\FT= 1\,\mathrm{N}\,,$$
which appears to be standard for commercially available spooling machines.
In a first very rough estimate the lateral pressures $\hat \P$ and $\check \P$, for the hexagonal case, are taken to be the same as $\bar\P$ from the quadrilateral case, as given in \eqref{eq: p_to_F_rel}.

Finally, the maximal pressure $\tilde \P$ for fibers at the bottom can be estimated by the weight of the spooled fiber $\sim 10\,\mathrm{kg}$ distributed over the area of a horizontal cross-section of the spool, resulting in
$$\tilde \P_\mathrm{max} =  1.3\,\mathrm{kPa}\,.$$
The actual value of $\tilde \P$ decreases with height along the spool, vanishing for the top layer. For the following numeric estimates we use this maximal value.

The GRAVITES estimates for the averaged differences of stresses are denoted by,
\begin{equation}
\delta \bar \sigma_{rr} : = \frac{1}{\pi a^2} 
 \int_U    \Big|\sigma_{rr}(r,\theta) \big|_{@1\,\mathrm{m}}- \sigma_{rr}(r,\theta)\big|_{@0\,\mathrm{m}} \Big|
 \, r\,
\dd r\, \dd \theta
\,,
 \label{12I24.11}
\end{equation}
the averaged  difference of displacements at the boundary are written as
\begin{equation}
\delta \bar u = \frac{1}{2 \pi} \int_{\partial U}     \| u \big|_{@1\,\mathrm{m}} -  u\big|_{@0\,\mathrm{m}}  \| \,\dd \theta
 \label{12I24.15}
\,,
\end{equation}
where $u=u_r e_r + u_\theta e_\theta$. 
We estimate the change in fiber length as
\begin{equation}
\delta L = L\Big[   - \frac{\poisson}{E}(\delta \bar \sigma_{rr} + \delta \bar \sigma_{\theta\theta} )   +
	 \alphat (T - T_0)\Big]
   \,,
\end{equation}
and we use
\begin{equation}
 \label{12I24.12}
  \delta \varphi= \beta\delta L
\end{equation}
to denote the change of phase of light due to the change of length of the waveguide.
The results  are summarised  in Table~\ref{tab:GRAVITES} for the single and double plane cases covered in Section~\ref{sec:step} and Section~\ref{sec:planesq}.
The configurations with four and six contact planes only give small corrections compared to the two plane case, since the additional pressures are independent of the height.
\begin{table}[h!]
$$
\begin{array}{||c||c|c|c|c||}
\hline
\hline
							&	\delta \grav^\text{single}  &	\delta \grav^\text{double}	&	\delta \mathfrak{p} = 10\,\mathrm{Pa}	&	\delta T = 10^{-2}\,\mathrm{K}
	 \\\hline
	 \delta \bar\sigma_{rr} [\mathrm{Pa}]	& 	2.5\cdot 10^{-7}		& 		3.4\cdot 10^{-7}		&	10 				& 0
	 \\\hline
	 \delta\bar u [\mathrm{\upmu m}]	&	8.3 \cdot 10^{-15}		&		9.6 \cdot 10^{-15}		&  8.4 \cdot 10^{-9} 		& 1.7 \cdot 10^{-7}
	 \\\hline
	 \delta L [\mathrm{\upmu m}]		&	1.2 \cdot10^{-7}	 	&		1.6 \cdot 10^{-7}		&  4.7  				& 180
	 \\\hline
	 \delta \varphi [\text{rad}]			&	7.0 \cdot 10^{-7}	 	&		9.4 \cdot 10^{-7}		&  28 					& 1080
	 \\\hline
	 \hline
\end{array}
$$
\caption{Estimates for the GRAVITES experiment described in \cite{HMMMCW}. The numbers correspond to a difference of height between the arms of 1 m. The columns isolate the effects with respect to their sources. We write $\delta \grav^\text{single}  $ for the effect calculated in the configuration of Section~\ref{sec:step} and $\delta \grav^\text{double}	$ for that in Section~\ref{sec:planesq}; $\delta \bar\sigma_{rr} $ is defined in \eqref{12I24.11}; $ \delta\bar u$ is the average transverse deformation of the waveguide as defined in \eqref{12I24.15}; $\delta \varphi$ is the change of phase of light due to the elastic elongation of the waveguide. }
\label{tab:GRAVITES}
\end{table}
This should be compared with the gravitational effect
\begin{equation}\label{12I24.1}
    \fbox{$
    \delta \varphi_\text{GRAVITES}=-6.52\times 10^{-5}\,\mathrm{rad}
    $}
\end{equation}
in the expected GRAVITES signal.

It has been proposed to place the interferometer in a centrifuge, with horizontal plane of rotation, with the arms of the interferometer placed at different distances from  the center of the centrifuge, achieving an acceleration gradient $\delta \grav \approx 10 \, \mathrm{m}/\mathrm{s}^2$ between the arms.
To estimate the difference of phase between the arms    we invoke the equivalence principle, modelling this problem by a radial gravitational field directed away from the axis of rotation of the centrifuge, with strength depending  upon the distance from the axis, and ignoring the Earth gravitational field which acts with constant strength normal to the plane of rotation. (Clearly a more precise model would be needed for the real experiment.) For future reference, the corresponding values are given in Table \ref{tab:GRAVITEScentri}; we do not include effects related to the change of temperature or pressure there, as such effects would depend strongly upon the experimental setup.
\begin{table}[h!]
$$
\begin{array}{||c||c|c||}
\hline
\hline
							&	\delta \grav^\text{single}  &	\delta \grav^\text{double}	
	 \\\hline
	 \delta \bar\sigma [\mathrm{Pa}]	& 	0.8				& 1.8					
	 \\\hline
	 \delta\bar u [\mathrm{\upmu m}]	&	2.3 \cdot 10^{-8}	 	&3.8\cdot 10^{-8}				
	 \\\hline
	 \delta L [\mathrm{\upmu m}]		&	0.4				& 0.8				
	 \\\hline
	 \delta \varphi	[\text{rad}] 			&	2.3			& 4.8 					
	 \\\hline
	 \hline
\end{array}
$$
\caption{Estimates for a centrifuge experiment with $\delta \grav =  10
\, \mathrm{m}/\mathrm{s}^2 $ at one meter separation. Notation as in Table~\ref{tab:GRAVITES}.}
\label{tab:GRAVITEScentri}
\end{table}

\appendix

\section{Relation between $\bar{\mathcal{P}}$ and $\alpham$}
\label{App24XI23.1}

The pressure terms $\bar \P$
 in the quadrilateral configuration, as well as $\bar \P$ and $\check \P$ in the hexagonal configuration are linked to the spooling tension.
We make this relation explicit for the case of a waveguide pressing against a rigid cylinder of radius $R$.
\begin{figure}[h]
 \centering
 \includegraphics[scale=1]{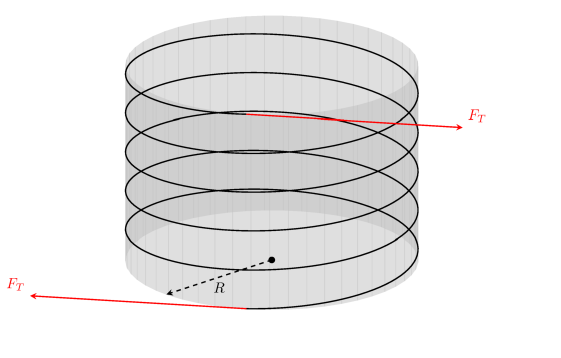}
 \caption{A waveguide wound around a rigid cylinder with radius $R$. We show a single layer of waveguide wrapped
 around the spool. Equal tension forces are applied at both the starting and end points so that the spool is in equilibrium. The spacing between the neighboring sections of the fiber has been increased for visual clarity.}
 \label{fig:sketch}
\end{figure}

We assume a static configuration, and
we ignore boundary effects at the point where
the contact between the waveguide and the spool is lost.
We consider   a waveguide, consisting of an optical fiber, which is wound $N$ times in one layer around a rigid cylinder of radius $R$.
We suppose that the contact interface of the waveguide and the cylinder has constant width $w$, and that the pressure, which we denote by $\phere$, exerted on the cylinder by the waveguide is constant across the whole area of contact. We assume that the axis of the cylinder is vertical, and we are interested in the pressure felt by the waveguide at the contact interface between the waveguide and the cylinder.

In a real-life situation there would be several layers of the waveguide, with various radii, with only the lowest one in contact with the cylinder, and the further ones in contact with neighbouring layers. Here we only consider the first layer, which is in contact with the cylinder. Similar considerations can be applied to describe  successive layers;
 one would then also need  to take into account the fact that the neighbouring layers are elastic and not rigid, as well as deformations of the neighbouring layers
arising from the Poisson ratio. Our analysis in the main body of the paper  sets the ground for further such investigations, which are left to future work.

\begin{figure}
  \centering
  \includegraphics[scale=1]{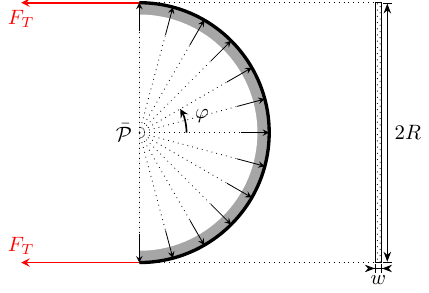}
  \caption{Free-body diagram of half of a single turn of waveguide around the spool  (this corresponds to a view from above in Figures~\ref{fig:coilquad} or~\ref{fig:sketch}, so that the vertical is orthogonal to the plane of the picture). The rectangle on the right-hand side shows the contact area with the width $w$  as viewed from the left. The arrows illustrate the direction of the (uniform) reaction force.}
  \label{fig: crosssection}
\end{figure}

We relate the pressure $\bar \P$ of Figure~\ref{fig:coilquad} to the tension force $\FT$ by considering the free-body diagram of a half of a single turn of a waveguide, i.e., we consider the intersection of the spool together with the waveguide and a plane on which the axis of the spool lies. The left Figure~\ref{fig: crosssection} shows the intersection when viewed from above the spool of Figure~\ref{fig:sketch}.
% fig: cross section of coil and waveguide

Let us imagine that we have a constant pressure $\bar \P$ acting from
the side as in Figure~\ref{fig:coilquad}, generating a force orthogonal  to the contact region   with the spool. This pressure generates a force $\mathcal{F}$ which should equate the tension forces, i.e., $\mathcal{F} = 2 \FT$, where
\begin{equation}
  \mathcal{F}
  :=
  \int_{-\pi/2}^{\pi/2} \bar \P \cos \varphi \, \otherA \, \dd \varphi
  =
  2 R w
  \int_{-\pi/2}^{\pi/2} \bar \P \cos \varphi \, \dd \varphi
\end{equation}
is the normal force due to the pressure $\bar \P$ acting on the contact area $\otherA$ and $\varphi$ is the angle shown in Figure~\ref{fig: crosssection} cover 180 degrees of a semicircle. Hence, $\mathcal{F}$ is the effective force caused by the pressure $\bar \P$.

We assume that the tension force is constant along the waveguide, and it is not affected during spooling. Therefore, we find $\mathcal{F} = 4 R w \bar \P$, which after equating the forces yields
\begin{equation}\label{eq: p_to_F_rel}
  \bar \P
  =
  \frac{\FT}{2 R w}
\,.
\end{equation}
Recall that the constant $\FHertz$ appearing in the Hertz constant problem is defined as the total force per unit length pushing the bodies into each other, which in our context translates to
$
\bar \P = \FHertz / w
$
.
Comparing with \eqref{eq: p_to_F_rel} we find
\begin{equation}\label{1XII24.31}
  \FHertz = \frac{\FT}{2R}
  \,.
\end{equation}
(Note that
this is independent of the number of windings in the spool, and that a virtual work calculation gives the same result.)
This value of $\FHertz$ can now be used  in Section~\ref{sec:Hertz} to calculate   $w$  by solving the  contact problem.

This formula  is valid for the first layer, as counted starting from the core of the spool.
One can generalize the result for the case where we have more layers that exert forces upon each other by writing
\begin{equation}\label{eq: pleft_pright_to_FT}
  \bar \P_{n} - \bar \P_{n+1}
  =
  \frac{\FT}{2 [R + (n-1) a] w_n}
\,,
\end{equation}
where
 $w_n$ is the contact width at the $n$'th layer, while $\bar \P_{n}$  and $\bar \P_{n+1}$  are the contact pressures there.

\section{Extending to non-symmetric solutions}
\label{app:non-symmetric}

Although the mirror symmetry assumption made in section \ref{sec:Elasticity} is well-motivated for our purposes we nevertheless give the full solution for completeness, demanding only regularity at $r = 0$ and $2\pi$-periodicity in $\theta$.

Discarding only terms incompatible with regularity and $2\pi$-periodicity in \eqref{eq:Michell}, as well as $B_0$, $D_1$ and $d_1$ which  do not contribute to the stresses, we have
\begin{align}\label{eq:Michell_adapted_non_sym}
	 \fsPhi (r,\theta)=\,&  D_0 r^2
		+ C_1 r^3 \cos (\theta)
		+ c_1 r^3 \sin (\theta)
	\\	
\notag	&+ \sum_{n\geq 2} \left[ \left( A_n r^n +  C_n r^{n + 2} \right) \cos(n \theta) +  \left( a_n r^n +  c_n r^{n + 2} \right) \sin(n \theta) \right]\,.
\end{align}
Again, the displacement can be derived by integrating \eqref{eq:StrainDisplacementrr}--\eqref{eq:StrainDisplacementrt}, yielding
\begin{align}
\notag	
	u_r (r, \theta)
	=\,&
	\frac{1}{2 \shearmod}
		\Bigg\{ 2 (1 - 2 \nu) D_0 r
		-
		\half \grav \rho (1 - 2 \nu) r^2 \cos (\theta)
		+ (1 - 4 \nu) r^2 (C_1 \cos (\theta) + c_1 \sin (\theta) )
		\\\notag
		&+ \sum_{n \geq 2} \Big( \left[- n A_n r^{n-1} + (2 - 4\nu - n) C_n r^{n+1} \right] \cos (n \theta)
		\\\notag
&+ \left[- n a_n r^{n-1} + (2 - 4\nu - n) c_n r^{n+1} \right] \sin (n \theta) \Big)
		\Bigg\}
		\\
&-
	\Xi \cos (\theta)
	 +
	\Xi_2 \sin (\theta)
	-
	\poisson \alpham  r
+
	(1 + \poisson )\alphat (T -T_0) r
\,,
	\label{eq:displacement_admissible_ur_non_sym}
	\\
	u_\theta (r, \theta)	
\notag	
	=\,&
	\frac{1}{2 \shearmod}
	\Big\{
		(5 - 4 \nu) r^2( C_1 \sin (\theta) - c_1 \cos (\theta))
		-
		\half \grav \rho (1 - 2 \nu) r^2 \sin (\theta)
\\\notag
	& +
	\sum_{n \geq 2} \Big( \left[ n A_n r^{n-1} + (4 - 4\nu + n) C_n r^{n+1} \right] \sin (n \theta)
	\\\notag
	&+  \left[ - n a_n r^{n-1} - (4 - 4\nu + n) c_n r^{n+1} \right] \cos (n \theta)  \Big)
	\Big\}
\\
&+
	\Xi \sin (\theta)
	+
	\Xi_2 \cos (\theta)
	+ \Dconst r
\,,
\label{eq:displacement_admissible_ut_non_sym}
\end{align}
with the integration constants determined by the boundary conditions $u_r (a,0) = 0 = u_\theta (a, 0)$ as above. Thus, identically
\begin{equation}
\begin{aligned}
	\Xi
	&=
	\frac{1}{2 \shearmod}
	\Big\{
		2 (1 - 2 \nu) D_0 a + (1 - 4 \nu) C_1 a^2
		-
		\half \grav \rho (1 - 2 \nu) a^2
\\
& \qquad\quad
		+
		\sum_{n \geq 2} \left[- n A_n a^{n-1} + (2 - 4\nu - n) C_n a^{n+1} \right]
	\Big\}
	-
	\poisson \alpham  a
	+
	(1 + \poisson )\alphat (T -T_0) a
\,,
\end{aligned}
 \label{24XI23.p1_non_sym}
\end{equation}
and further
\begin{equation}
\begin{aligned}
	\Xi_2 	
	&=
	\frac{1}{2 \shearmod}
	\Big\{
		(5 - 4 \nu) a^2  c_1  +
	\sum_{n \geq 2}
 \left[  n a_n a^{n-1} + (4 - 4\nu + n) c_n a^{n+1} \right]
	\Big\}
	- \Dconst a
\,,
\end{aligned}
 \label{24XI23.p2_non_sym}
\end{equation}
provided that the sums converge.

Continuing to the boundary conditions, we immediately restrict to the frictionless case, taking
\begin{equation}
	\sigma_{r\theta} |_{\p U} = 0\,.
	\label{eq:28XI23.3}
\end{equation}
The boundary condition for $\sigma_{rr}$ can now be a general Fourier series which we write in the suggestive form
\begin{align}
	\t{\sigma}{_r_r}|_{\p U} = f(\theta) =  f_0 + \sum_{n \geq 1} \left[(f_{-n} + f_n) \cos(n \theta) + i (f_{-n} - f_n) \sin(n \theta) \right]\,,
	\label{eq:cosexp_non_sym}
\end{align}
with
\begin{equation}
	f_n = \frac{1}{2\pi} \int_{-\pi}^\pi \dd \theta\; f(\theta) e^{i n \theta}\,.
\end{equation}

These boundary conditions determine the coefficients in \eqref{eq:Michell_adapted_non_sym} algebraically by comparing to the corresponding stresses \eqref{eq:stress_cyl_rr}--\eqref{eq:stress_cyl_rt}.
We find
\begin{equation}
	C_1 = c_1 = 0 \,, \  C_n = \frac{1 - n}{a^2 (1+n)} A_n \ \mbox{and} \  c_n = \frac{1 - n}{a^2 (1+n)} a_n \,,
\end{equation}
via \eqref{eq:28XI23.3}, leading to
\begin{equation}
	\sigma_{rr}|_{\p U} = 2 D_0 - a \grav \rho \cos(\theta) + \sum_{n\geq 2} 2 (1-n)   a^{n-2}  \left[ \sin ( n \theta) a_n +  \cos ( n \theta ) A_n \right]\,,
\end{equation}
and comparing with \eqref{eq:cosexp_non_sym} for the remaining coefficients
\begin{align}
	D_0 &= \frac 12 f_0\,,
		\\
	A_n &= -\frac{1}{2 (n-1)} a^{2-n} (f_{-n} + f_n)\,,
			\\
	a_n &= -\frac{i}{2 (n-1)} a^{2-n}  (f_{-n} - f_n)\,,
\end{align}
for $n \geq 2$, as well as
\begin{align}
	f_{-1} + f_1 &= - a \grav \rho\,,
	\\
	f_{-1} - f_1 & = 0\,.
	\label{eq:28XI23.4}
\end{align}

Note that, similarly to the symmetric case the space of possible boundary conditions is constrained by the body force. Without any body forces oriented along the $x$-direction \eqref{eq:28XI23.4} requires that the $\sin (\theta)$ term in \eqref{eq:cosexp_non_sym} vanishes.

In the simplest case, considering line forces from the left and right and a line support from the bottom, i.e. boundary conditions
\begin{align}
	\t{\sigma}{_r_\theta}|_{\p U} &= 0\,,
		\\
	\t{\sigma}{_r_r}|_{\p U} &= \P \delta(\theta) + \hat \P \delta (\theta - \pi/2) + \check \P \delta (\theta + \pi/2)\,,
\end{align}
equation \eqref{eq:28XI23.4} implies $\hat \P = \check \P$, thus forcing mirror symmetry.

\bigskip

{\noindent \sc Acknowledgements:}
Useful discussions with Thomas Mieling and Philip Kornreich are acknowledged. We are very grateful to Arash Yavari for advice and comments about a previous version of the manuscript.
Research supported in part by the Austrian Science Fund (FWF), Project P34274 and by the European Union (ERC, GRAVITES, project no 101071779).
 H.B. was supported in part by the European Research Council (ERC) under the European Union's Horizon 2020 research and innovation program (grant agreement ERC advanced grant 740021--ARTHUS, PI: Thomas Buchert). 
Views and opinions expressed are however those of the authors only and do not necessarily reflect those of the European Union or the European Research Council Executive Agency. Neither the European Union nor the granting authority can be held responsible for them. 

\providecommand{\bysame}{\leavevmode\hbox to3em{\hrulefill}\thinspace}
\providecommand{\MR}{\relax\ifhmode\unskip\space\fi MR }
% \MRhref is called by the amsart/book/proc definition of \MR.
\providecommand{\MRhref}[2]{%
  \href{http://www.ams.org/mathscinet-getitem?mr=#1}{#2}
}
\providecommand{\href}[2]{#2}

\end{document}